\documentclass[twocolumn, tighten]{aastex631}

\usepackage[T1]{fontenc}
\usepackage{graphicx}	
\usepackage{amsmath}	
\usepackage{amssymb}	
\usepackage{hyperref}
\usepackage{makecell}
\usepackage{color} 		
\usepackage{comment}    

\newcommand{\radshock} {\texttt{radshock}}

\shortauthors{F. Alamaa}
\shorttitle{Intrapulse evolution in photospheric GRBs}

\begin{document}
\label{firstpage}
\title{Intrapulse spectral evolution in photospheric gamma-ray bursts}

\correspondingauthor{Filip Alamaa}
\email{filip.alamaa@iap.fr}

\author[0000-0001-7414-5884]{Filip Alamaa}
\affiliation{Institut d'Atrophysique de Paris, Sorbonne Universit\'e and CNRS, UMR 7095, 98 bis bd Arago, F-75014 Paris, France}
\affiliation{Department of Physics, KTH Royal Institute of Technology, and The Oskar Klein Centre, SE-10691 Stockholm, Sweden}

\begin{abstract}
Photons that decouple from a relativistic jet do so over a range of radii, which leads to a spreading in arrival times at the observer. 
Therefore, changes to the comoving photon distribution across the decoupling zone are encoded in the emitted signal. 
In this paper, we study such spectral evolution occurring across a pulse. We track the radiation from the deep subphotospheric regions all the way to the observed time-resolved signal, accounting for emission at various angles and radii. We assume a simple power-law photon spectrum injection over a range of optical depths and let the photons interact with the local plasma. 
At high optical depths, we find that the radiation exists in one of three characteristic regimes, two of which exhibit a high-energy power law. 
Depending on the nature of the injection, this power law can persist to low optical depths and manifest itself during the rise time of the pulse with a spectral index $\beta \approx \alpha - 1$, where $\alpha$ is the low-energy spectral index. 
The results are given in the context of a gamma-ray burst jet but are general to optically thick, relativistic outflows.  
\end{abstract}

\section{Introduction}
Gamma-ray burst (GRB) jets are initially optically thick. As the jet expands, the density drops and any trapped radiation can start leaking out towards the observer. The distance from the central engine where the ejecta transitions from optically thick to optically thin is called the photospheric radius, $R_{\rm ph}$, and the released radiation is called photospheric emission. Photospheric emission is an inevitable part of the fireball model and many magnetic jet acceleration models \citep{CavalloRees1978, Paczynski1986, Goodman1986, Piran1999, SpruitDaigneDrenkhahn2001}, and its role in the GRB prompt observations has been studied extensively \citep{DaigneMochkovitch2002, Ryde2005, ReesMeszaros2005, Giannios2006}.

Photospheric emission is inherently probabilistic in nature since it is related to photons experiencing their last scattering. Assuming a spherically symmetric outflow, \citet{Abramowicz1991} showed that the optical depth towards the observer depends on the angle between the radial direction and the observer line-of-sight, $\theta$. \citet{Peer2008} and \citet{Beloborodov2011} constructed a probability function for a photons last scatting as a function of $\theta$ and radius from the central engine, $r$. They found that photons decouple across a wide range of optical depths. \citet{Lundman2013} extended upon these works by relaxing the assumption of spherical symmetry, considering the effect of jet structure and arbitrary viewing angle on the observed spectrum. 

Photons that decouple from high optical depths outrun the plasma and are the first to arrive at the observer. They are followed by photons that experienced their last scattering at progressively lower optical depths. The spread in arrival times leads to a short-duration pulse in the observer frame. Time-resolved emission across such a pulse was considered in spherical symmetry by \citet{PeerRyde2011}. Time-resolved emission from the photosphere within a structured jet was studied by \citet{Meng2019}. In addition, \citet{Meng2019} also modeled a long-lasting and varying central engine by letting the bulk Lorentz factor and luminosity evolve with time. 

All of the aforementioned works have considered the comoving photon distribution to be in thermodynamic equilibrium with the plasma. 
In this case, the radiation is effected only by adiabatic cooling as the jet propagates outwards. This effect, together with the Doppler factor being angle dependent, lead to a broadening of the observed time-integrated spectrum, which consists of a superposition of many Doppler boosted comoving spectra emitted at various radii and angles to the line-of-sight \citep{Beloborodov2010}. 

However, a GRB jet is a highly chaotic system and energy dissipation below the photosphere is expected \citep{ReesMeszaros2005}. The dissipation may be due to radiation-mediated shocks \citep{LevinsonBromberg2008, Bromberg2011b, Levinson2012, Beloborodov2017, Samuelsson2022, SamuelssonRyde2023}, magnetic reconnection \citep{SpruitDaigneDrenkhahn2001, Drenkhahn2002, Giannios2006}, turbulence \citep{Zrake2019}, shearing flows in a structured jet \citep{Ito2013, VyasPeer2023}, or nuclear collisions between protons and neutrons \citep{Beloborodov2010}. Dissipation destroys the thermodynamic equilibrium and distorts the photon distribution, which may appear highly non-thermal. When the dissipation ceases, the photon distribution will start to re-establish a new equilibrium. 
Accounting for dissipation, the spectral change across the decoupling region can be much more dramatic compared to the scenario with adiabatic cooling only. 


In this paper, we study the time-resolved signal accounting for dissipation and thermalization. In contrast to \citet{Meng2019}, we do not model the central engine variation. Instead, we focus on a thin slice of the jet with an observed width $\delta r \sim r/\Gamma^2$, where $\Gamma$ is the jet bulk Lorentz factor. Therefore, the observed spectral evolution presented in this work is solely due to changes of the photon distribution within the jet. 
A complex GRB light curve would consist of radiation from many such slices, where each slice would create a subpulse with its own evolution. 

The dissipation is modeled by simply injecting a spectral power-law distribution of photons across a range of optical depths. We show that under these conditions, the radiation in the deep subphotospheric region exists in one of three characteristic regimes, which we call the slow, the marginally fast, and the fast Compton regime. These regimes have a direct analog in the slow and fast synchrotron cooling regimes, hence the chosen names. Each regime has a typical spectral shape and we characterize the relation between the spectral indices in each case. 
Tracking the photon distribution as a function of optical depth, we generate the time-resolved signal in the observer frame accounting for contributions from different angles and radii. We show that dissipation at low optical depths can lead to a significant intrapulse spectral evolution in the observer frame. 


Throughout this paper, we use the term thermalization to indicate that the radiation tends towards a kinetic equilibrium when interacting with the plasma, with high-energy photons losing energy and low-energy photons gaining energy. Furthermore, the terms downscattering and upscattering are used to describe photons losing or gaining energy in a scattering event, respectively. 

The paper is structured as follows. In Section \ref{sec:general_picture}, we introduce the general picture. We obtain the radiated energy as a function of optical depth and describe the model used in the paper. In Sections \ref{sec:subphotospheric_evolution} to \ref{Sec:results}, we follow the evolution of the photon distribution on its journey, from the deep subphotospheric regions (Section \ref{sec:subphotospheric_evolution}), across the decoupling zone (Section \ref{sec:evolution_across_decoupling_zone}), to the observed signal (Section \ref{Sec:results}). We discuss our results, with a specific emphasis on assumptions, in Section \ref{Sec:discussion} and we conclude in Section \ref{Sec:conclusion}.

\section{General picture}\label{sec:general_picture}
\subsection{Radiated energy as function of optical depth}
The radiated energy as a function of optical depth is a combination of the probability of last scattering at a given angle and optical depth, as well as the comoving radiation energy density at that point. When there is no energy dissipation, only two things are needed obtain the radiated energy as a function of optical depth: 1) a photon decoupling probability function and 2) a cooling function that describes the energy losses due to adiabatic expansion.

Throughout the paper, we assume the outflow to be spherically symmetric, i.e., that we are observing the jet on-axis and that the jet properties do not vary with angle within $\sim 1/\Gamma$ to the line-of-sight. 
In the case of spherical symmetry, the probability function for a photon to decouple as a function of optical depth and angle was derived using radiative transfer in the ultra-relativistic regime by \citet{Beloborodov2011}. In this work, we adopt the expression from Appendix C in \citet{SamuelssonRyde2023}, which is identical to the expression given in \citet{Beloborodov2011} but rewritten as a function of optical depth, $\tau$. It reads 
\begin{equation}\label{eq:f}
\begin{split}
    f(\tau, \mu') =& \frac{1}{4}\left\{ \frac{3}{2} + \frac{1}{\pi} \arctan \left[ \frac{1}{3} \left( \tau - \tau^{-1} \right)\right]\right\} \, \\
    &\times \exp \left[-\frac{\tau}{6}\left(3 + \frac{1 - \mu'}{1 + \mu'}\right)\right].
\end{split}
\end{equation}
\noindent Here, $\mu' = \cos(\theta')$, where $\theta'$ is the angle between the radial direction and the line-of-sight as measured in the frame comoving with the outflow. 

The optical depth in equation \eqref{eq:f} is measured for a fluid element \textit{along the radial direction} towards infinity.
If the electron density decreases with radius as $r^{-2}$, as expected in the GRB coasting phase, then $\tau$ is related to radius as \citep[e.g.,][]{Beloborodov2011}
\begin{equation}\label{eq:tau}
    \tau = \frac{R_{\rm ph}}{r}, \qquad R_{\rm ph} = \frac{L \sigma_{\rm T}}{4 \pi m_p c^3 \Gamma^3},
\end{equation}
\noindent where $L$ is the isotropic equivalent luminosity and $\sigma_{\rm T}$ is the Thomson cross section. 
Equation \eqref{eq:tau} assumes a negligible amount of pairs at the photosphere and that all magnetic energy initially present in the jet has been converted into kinetic energy below the photosphere.

Under the assumption of no dissipation, the total energy of the comoving radiation in the subphotospheric region of the jet is only affected by adiabatic cooling. \citet{SamuelssonRyde2023} found a ``photon cooling function'', $\phi(\tau)$, that well described the energy loss at high and low optical depths, including the transition across the photosphere. The cooling function is given by
\begin{equation}\label{eq:phi}
    \phi(\tau) = \frac{\tau^{2/3} + 0.2}{1.2},
\end{equation}
\noindent which was found to give very good agreement with simulations. The average photon energy at arbitrary optical depth, ${\bar \epsilon}(\tau)$, is related to the average photon energy at the photosphere, ${\bar \epsilon}_{\rm ph}$, as ${\bar \epsilon}(\tau) = \phi(\tau){\bar \epsilon}_{\rm ph}$. At high optical depths, $\phi(\tau)$ corresponds to an ideal adiabatic cooling of the photon distribution as $\propto {\bar r}^{-2/3}$, expected in the coasting phase of GRBs from conservation of entropy \citep[e.g.,][]{Peer2015}. At optical depths $\tau \ll 1$, all photons are free-streaming and $\phi(\tau)$ is constant. 

\begin{figure}
    \centering
    \includegraphics[width=\linewidth]{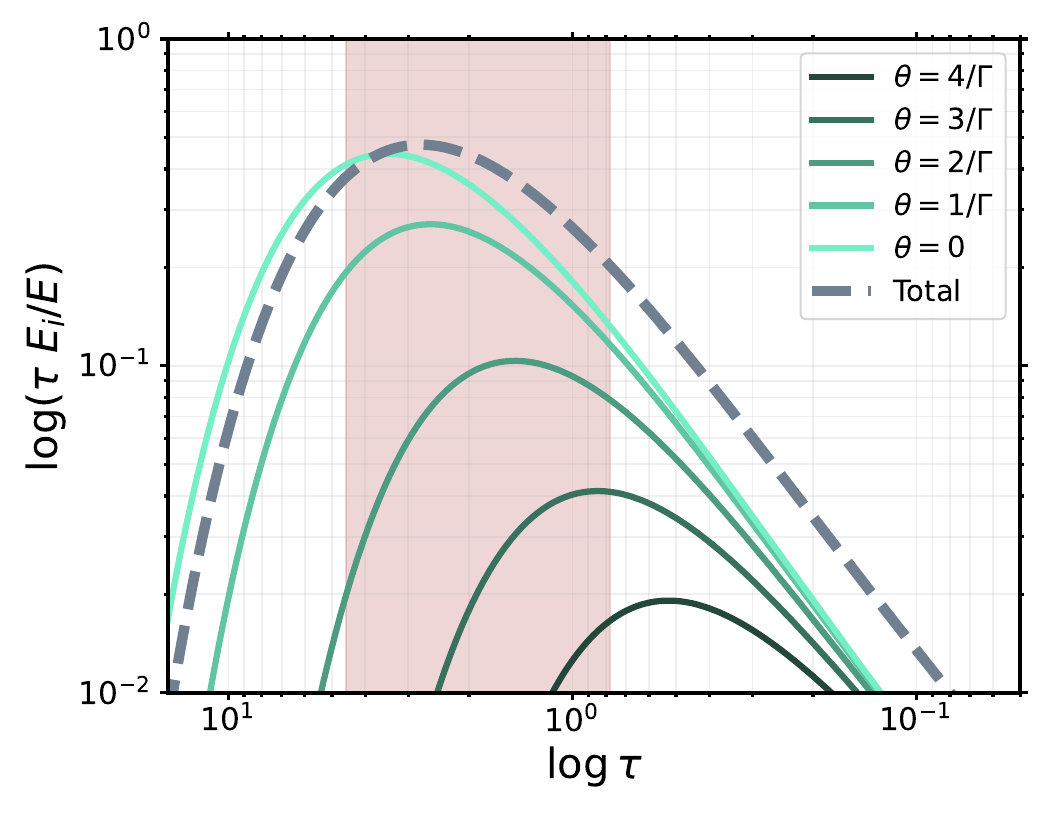}
    \caption{Normalized radiated energy as a function of optical depth and angle in a spherically symmetric outflow. The different solid lines show emission at different angles compared to the line-of-sight calculated using equation \eqref{eq:E_tau_mu}, and the grey, dashed line shows the angle-integrated radiated energy calculated using equation \eqref{eq:E_tau}. The quantity $E_i$ indicates $dE/d\tau d\mu'$ and $dE/d\tau$ for the solid lines and dashed line, respectively. The shaded region shows where $2/3$ of the energy is radiated. One can see that photons are emitted over a wide range of optical depths, and that the radiated energy peaks at $\tau \sim 3$. The figure is generated with $\Gamma = 100$ but is essentially identical as long as $\Gamma > 10$. A similar figure for the photon number can be found in \citet{Beloborodov2011}.}
    \label{fig:energy_vs_tau}
\end{figure}

With the photon cooling function, one can calculate the normalized radiated energy as a function of $\tau$ and $\mu'$ in a spherically symmetric outflow as
\begin{equation}\label{eq:E_tau_mu}
    \frac{E_{\tau,\mu'}}{E} = \frac{f(\tau, \mu') \, \phi(\tau)}{\iint f \, \phi\, d\tau d\mu'}. 
\end{equation}
\noindent Above, $E_{\tau,\mu'} = dE/d\tau d\mu'$ and $E = \iint E_{\tau,\mu'}d\tau d\mu'$. This quantity, multiplied by $\tau$, is plotted in solid green lines in Figure \ref{fig:energy_vs_tau} for different observing angles, $\theta$. The angle $\theta$ is related to $\mu'$ via $\mu' = (\mu - \beta)/(1-\mu\beta)$, where $\mu = \cos(\theta)$ and $\beta = \sqrt{1-1/\Gamma^2}$. The multiplication with $\tau$ ensures that the curve height indicates where most energy is radiated. 

The angle-integrated normalized radiated energy can be calculated as
\begin{equation}\label{eq:E_tau}
    \frac{E_{\tau}}{E} = \frac{\int f(\tau, \mu') \, \phi(\tau) d\mu'}{\iint f \, \phi\, d\tau d\mu'},
\end{equation}
\noindent where, $E_\tau = dE/d\tau$. This, again multiplied by $\tau$, is shown by a dashed grey line in Figure \ref{fig:energy_vs_tau}. Since scatterings do not change the energy in the photon bath if the electrons are kept at the Compton temperature, Figure \ref{fig:energy_vs_tau} is always accurate as long as dissipation is absent. 

It is clear from the figure that photons start free streaming at different depths. 
Additionally, photons reaching the observer from larger angles to the line-of-sight are preferentially emitted further out and contribute less to the total observed emission. Both these results are in agreement with previous works \citep{Peer2008, Beloborodov2011, Begue2013}. It is interesting to note that the radiated energy peaks already at $\tau \sim 3$.

\subsection{Jet profile}\label{sec:jet_profile}
%
%

\begin{figure}
\centering
    \includegraphics[width=\linewidth]{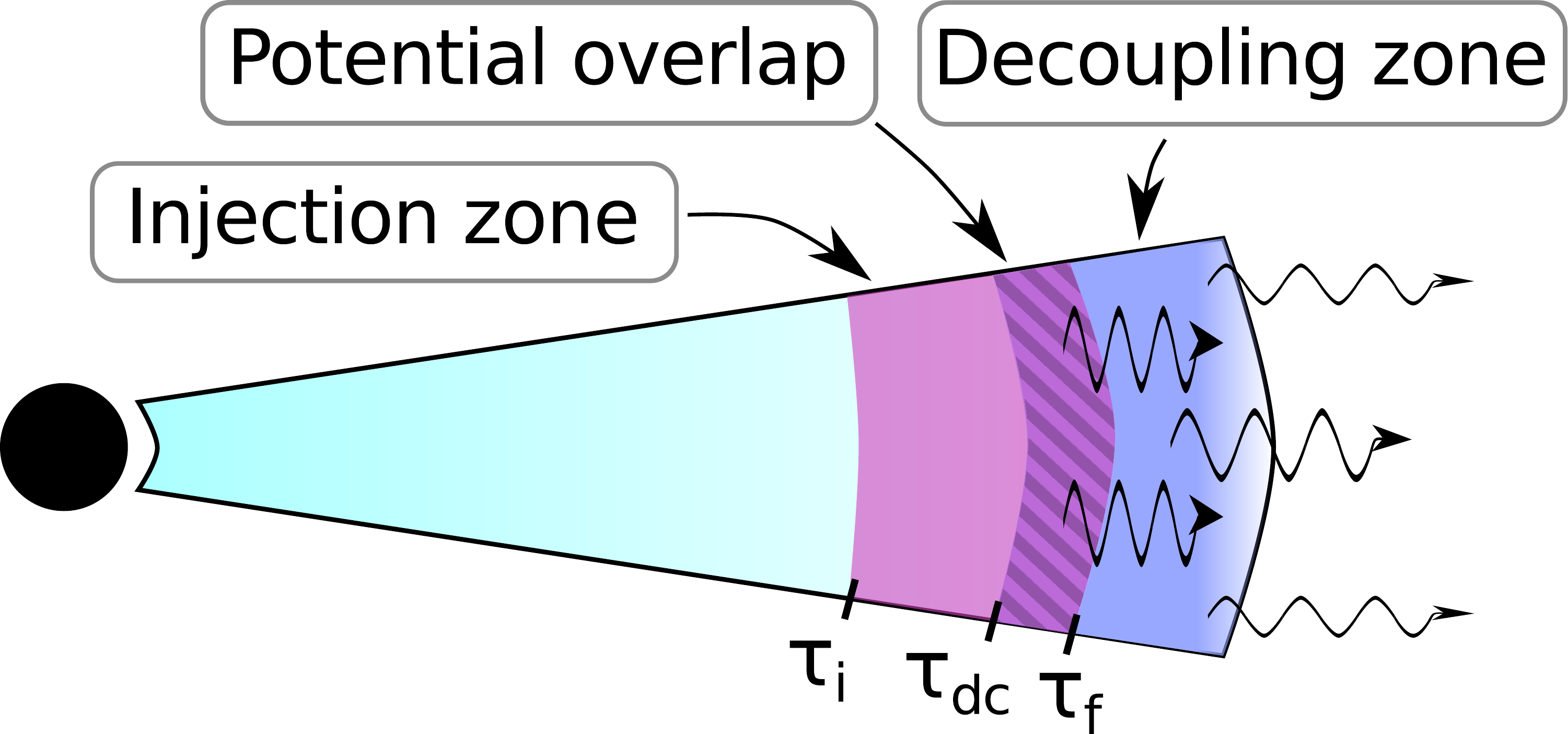}
    \caption{Schematic showing the scenario envisioned in this paper. The plasma moves from left to right, passing through each zone. In the injection zone between $\tau_i$ and $\tau_f$, some dissipation event continuously injects a power-law distribution of photons. The injected radiation interacts with the local plasma. Photons start leaking out of the plasma in the decoupling zone, at $\tau < \tau_{\rm dc}$. If injection occurs within, or close to, the decoupling zone, the observed spectral shape may change significantly with time.}
    \label{fig:jet_scematic}
\end{figure}

The scenario envisioned in the current paper is given in the schematic in Figure \ref{fig:jet_scematic}. We assume that some dissipation process generates a power-law distribution of photons as a function of energy. The assumption of a power-law spectral distribution is discussed in more detail in Section \ref{sec:power_law_injection}. These photons are continuously injected into a region that we call the \textit{interaction region}. We assume a singe-zone approximation for the interaction region, implying that the whole region is causally connected. The injection continues throughout the injection zone (red), which spans an optical depth $\Delta \tau_{\rm inj} = \tau_i - \tau_f$. 

All injected photons are accumulated in the interaction region. Furthermore, the interaction region contains only the injected photons, i.e., there are no photons in the interaction region at optical depths $\tau > \tau_i$. This scenario mimics that of a region downstream of a subphotospheric shock occurring at $\tau_i$. In the case of a shock, thermal photons from the upstream are energized as they traverse the shock region, after which they are advected downstream. Thus, all photons in the downstream region have passed through the shock and the downstream itself did not exist before the shock was initiated. The photons in the interaction region interacts with the local plasma through thermal Comptonization. Meanwhile, the whole ejecta is moving outwards towards the photosphere (fluid elements move from left to right in the cartoon). 

Photons have a non-zero probability to decouple anywhere in the jet as evident from equation \eqref{eq:f}. However, $\sim 99$\% of the energy is radiated in the ``decoupling zone'' (blue), at optical depths $\tau \leq \tau_{\rm dc}$, where $\tau_{\rm dc} \approx 10$. 
The decoupling zone, therefore, spans an optical depth of $\Delta \tau_{\rm dc} \approx 10$. In an ultra-relativistic outflow, the average number of scatterings per photon travelling from $r$ to infinity is roughly equal to the optical depth at $r$. 
Thus, on average, photons experience several scatterings while traversing the decoupling zone. Note that we introduce the notion of a decoupling zone to facilitate the discussion. However, all results presented are obtained using the probability distribution in equation \eqref{eq:f}.


In the following sections, we will follow the evolution of the photon distribution in the interaction region as it traverses the GRB jet. 

\section{Spectral evolution across the injection zone}\label{sec:subphotospheric_evolution}
Many authors have made detailed calculations regarding the evolution of the subphotospheric photon distribution in the presence of various dissipation mechanisms \citep[e.g.,][]{Beloborodov2010, VurmBeloborodov2016, Parsotan2018, Samuelsson2022}. In this paper, we consider a simple one-zone model, which is continuously injected with photons in a power-law spectral distribution between the energies $\epsilon_{\rm min}$ and $\epsilon_{\rm max}$. Photon energies are measured in units of $m_e c^2$ throughout the paper. Furthermore, we assume that $\epsilon_{\rm max} \leq 1$, such that we can neglect relativistic effects in the comoving frame. The slope of the injected spectrum in photon number is $\alpha_{\rm inj}$, i.e., ${\mathcal N}_\epsilon \propto \epsilon^{\alpha_{\rm inj}}$, where ${\mathcal N}_{\epsilon}$ is the specific photon number density. We show that under these simplified assumptions, the photon distribution can exist in three characteristic regimes. In this section, we assume that the injection is always active. 

\subsection{Slow, fast, and marginally fast Compton regime}\label{sec:fast_slow_Compton}

In the subphotospheric region of a GRB jet, the photons vastly outnumber the electrons once the initial pairs have recombined \citep{Bromberg2011b}.\footnote{This would not be true in a cold, magnetically accelerated jet.} Under such conditions, the electron scattering time is much shorter compared to the photon scattering time. This implies that to a good approximation, the electrons are always kept in a thermal distribution at the local Compton temperature, $\Theta_{\rm C}$. The Compton temperature is defined as the temperature that satisfies no net energy exchange between the photon and electron populations. Note that due to the difference in scattering length, the photon distribution need not be thermal.

In Compton scattering between an isotropic photon field and thermal electrons, the average relative energy change per scattering is given by \citep{RybickiLightman1979}
\begin{equation}\label{eq:delta_eps}
    \frac{\Delta \epsilon}{\epsilon} = 4\Theta_{\rm C} - \epsilon,
\end{equation}
\noindent where $\Delta \epsilon$ is the photon energy change in the scattering, $\epsilon$ is the photon energy, and $\Theta_{\rm C}$ is the electron temperature, all measured in units of electron rest mass energy $m_e c^2$. 
Equation \eqref{eq:delta_eps} is valid for photons with energies $\epsilon \ll 1$. It is clear that high-energy photons lose energy more quickly than low-energy photons gain energy. From the equation above, one can derive that all photons with energies $\epsilon > 1/N_{\rm sc}$ lose at least half of their energy after $N_{\rm sc}$ number of scatterings. As the number of scatterings is roughly equal to the optical depth traversed, all photons with energies $\epsilon > 1/\Delta \tau$ lose a significant amount of energy over an optical depth $\Delta \tau$. 
The downscattering process continues until $1/\Delta\tau$ becomes comparable to $4\Theta_{\rm C}$. 
This leads to three different possible regimes. 

\textit{Slow Compton regime: $4\Theta_{\rm C}\Delta \tau < 1$.} In this case, only the high-energy photons with energies satisfying $\epsilon > 1/\Delta \tau$ have time to be downscattered. No other photons are significantly effected, regardless of whether their energy is above or below $4\Theta_{\rm C}$. This leads to a spectrum with power-law slope $\alpha = \alpha_{\rm inj}$ between $\epsilon_{\rm min}$ and $1/\Delta \tau$. At energies higher than $1/\Delta \tau$, the power law steepens due to the additional Compton losses. Since the relative energy loss is proportional to the photon energy, the power-law slope becomes $\beta = \alpha_{\rm inj} - 1$.\footnote{This result can also be obtained by evaluating the steady-state solution to the Kompaneets equation \eqref{eq:kompaneets_spherical} in the region $4\Theta_{\rm C} \ll \epsilon \ll \epsilon_{\rm max}$, ignoring adiabatic cooling.} This result is valid only when $\alpha_{\rm inj} < -1$, above which no steady-state solution exists (see further discussion in Section \ref{sec:alpha_beta_relation}). A special case of the slow Compton regime is when $\epsilon_{\rm max}< 1/\Delta \tau$, in which case the spectral shape is simply similar to that of the injected spectrum.

The shape of the photon distribution in this regime is completely analogous to the slow cooling distribution of charged particles downstream of a collisionless shock, hence the chosen name. In the synchrotron case, charged particles are injected as a power law with index $-p$ above the injection Lorentz factor $\gamma_m$. In the slow cooling regime, particles below the cooling Lorentz factor $\gamma_c>\gamma_m$ do not lose energy efficiently and, thus, the particle distribution between $\gamma_m$ and $\gamma_c$ has slope $-p$. Above $\gamma_c$, particles are cooled due to synchrotron losses and the energy loss timescale is inversely proportional to the particle energy. This steepens the power-law slope above $\gamma_c$ to $-p-1$ \citep{Sari1998}. The same thing occurs here, excepts that it is the photons that are the primary particle population. The break occurs at $\epsilon_c = 1/\Delta \tau$.

\textit{Fast Compton regime: $4\Theta_{\rm C}\Delta \tau > 1$.} In this case, all high-energy photons have time to downscatter, leading to a pile up around $4\Theta_{\rm C}$. Furthermore, the Compton $y$-parameter for the low-energy photons, $y \equiv 4\Theta_{\rm C}\Delta \tau$, is larger than unity. This implies that all low-energy photons will significantly increase their energy, leading to a hardening of the low-energy part. If the difference between $4\Theta_{\rm C}$ and $\epsilon_{\rm max}$ is large enough, a power-law segment with $\beta = \alpha_{\rm inj} - 1$ develops in the region $4\Theta_{\rm C} \ll \epsilon \ll \epsilon_{\rm max}$. The analogy between this scenario and the fast cooling synchrotron scenario is less clear here, since some photons gain energy in the fast Compton regime while all charged particles in the synchrotron fast cooling regime lose energy. 

\textit{Marginally fast Compton regime: $4\Theta_{\rm C}\Delta \tau \sim 1$.} In the intermediate regime, we have $y\sim 1$ meaning that the low-energy photons are marginally affected. Furthermore, a significant pile up of high-energy photons is not yet visible around $4\Theta_{\rm C}$. In this case, the spectrum consists of a power law with slope $\alpha \sim \alpha_{\rm inj}$ between $\epsilon_{\rm min}$ and $4\Theta_{\rm C}\sim 1/\Delta \tau$, a possible minor pile-up above $4\Theta_{\rm C}$, and a power law with slope $\beta = \alpha_{\rm inj} - 1$ between $\gtrsim 4\Theta_{\rm C}$ and $\epsilon_{\rm max}$. 

\begin{figure*}
\centering
    \includegraphics[width=0.45\linewidth]{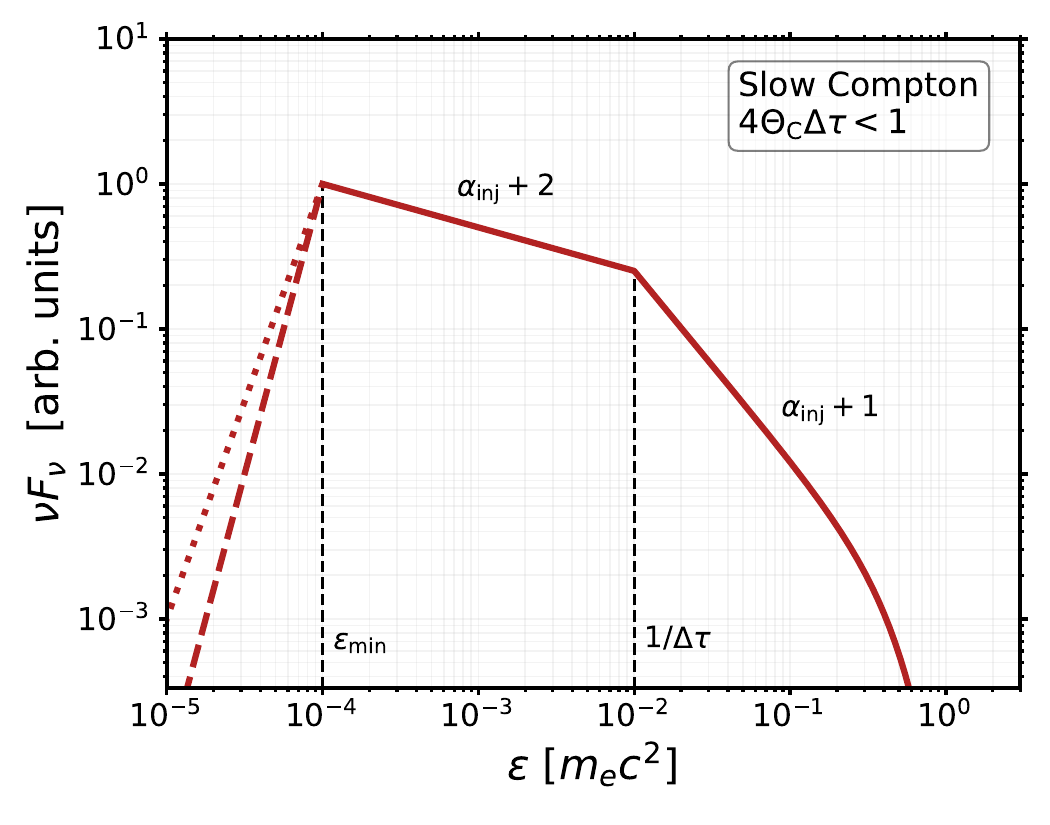}
    \includegraphics[width=0.45\linewidth]{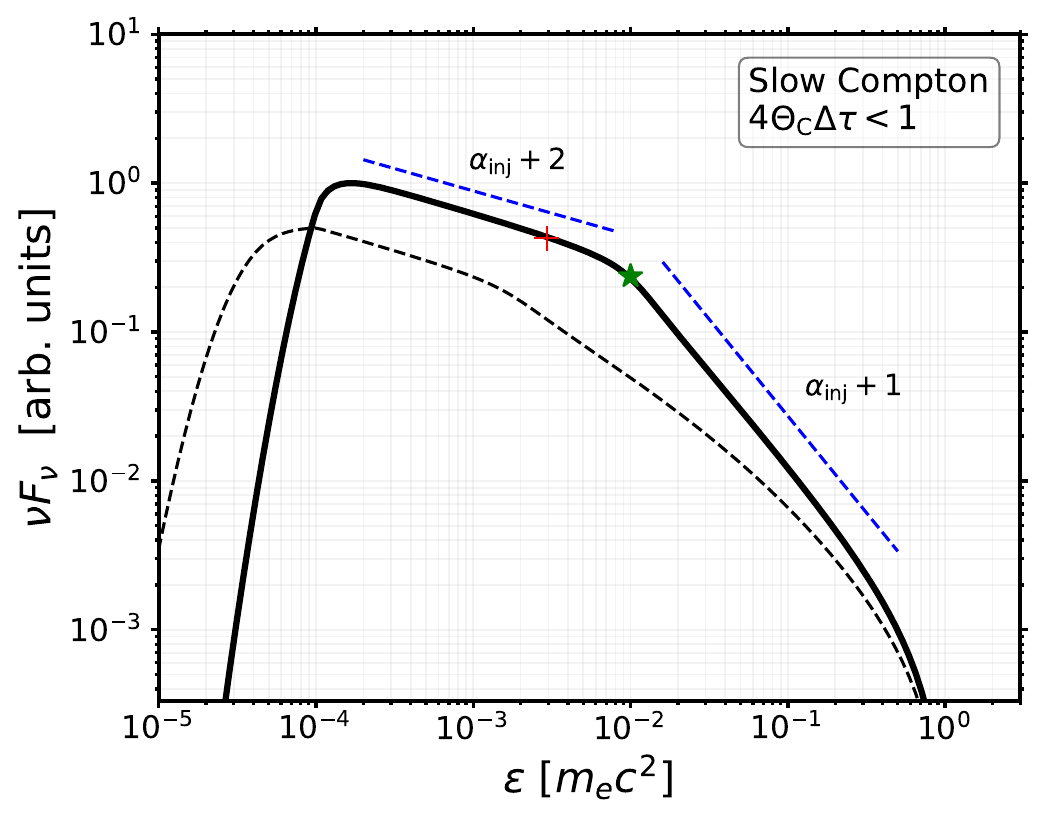}\\
    \includegraphics[width=0.45\linewidth]{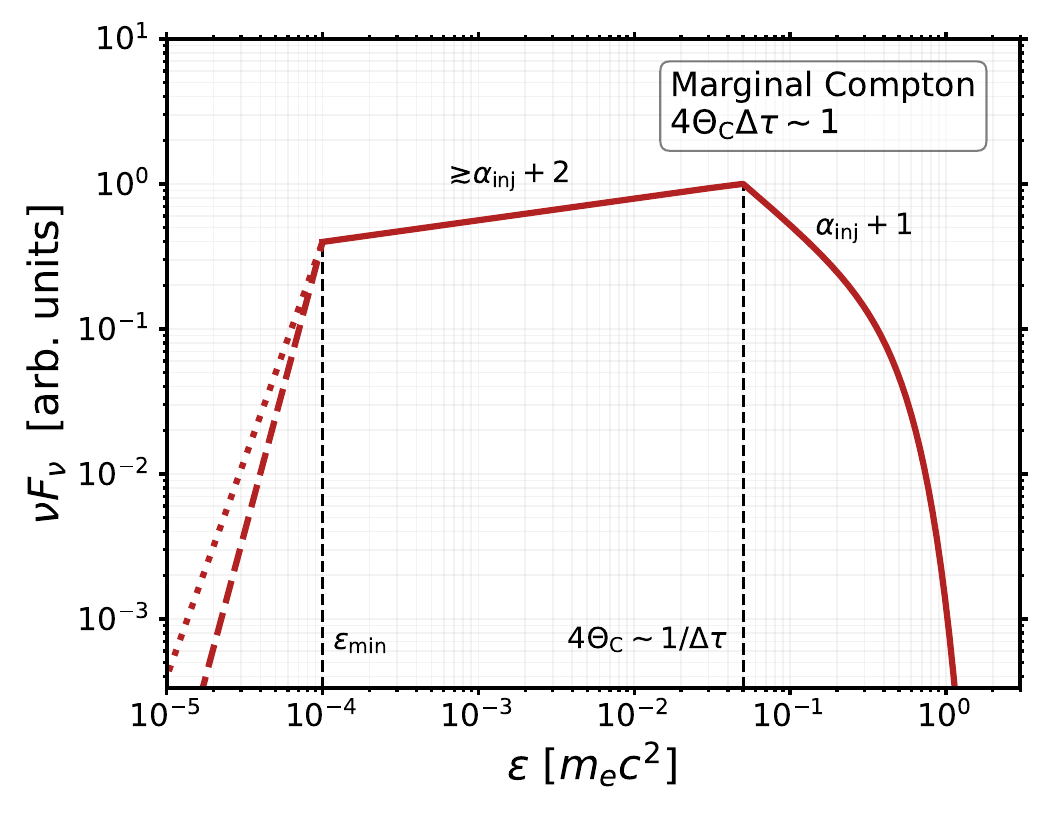}
    \includegraphics[width=0.45\linewidth]{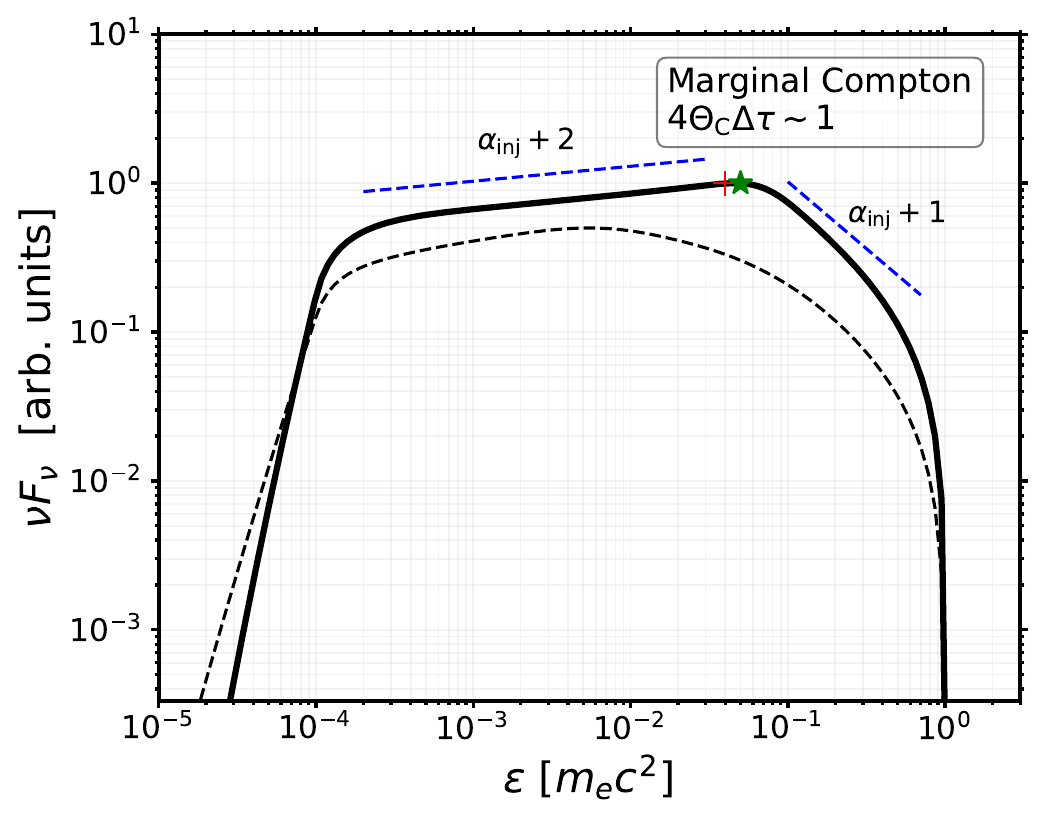}\\
    \includegraphics[width=0.45\linewidth]{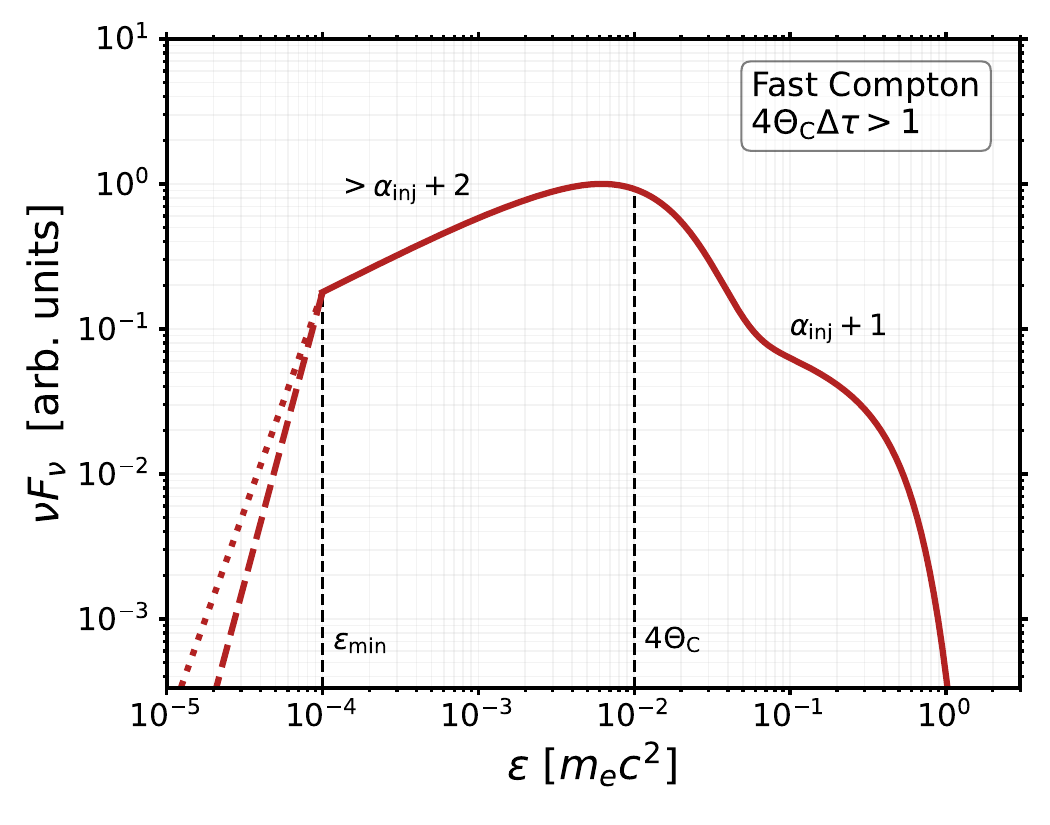}
    \includegraphics[width=0.45\linewidth]{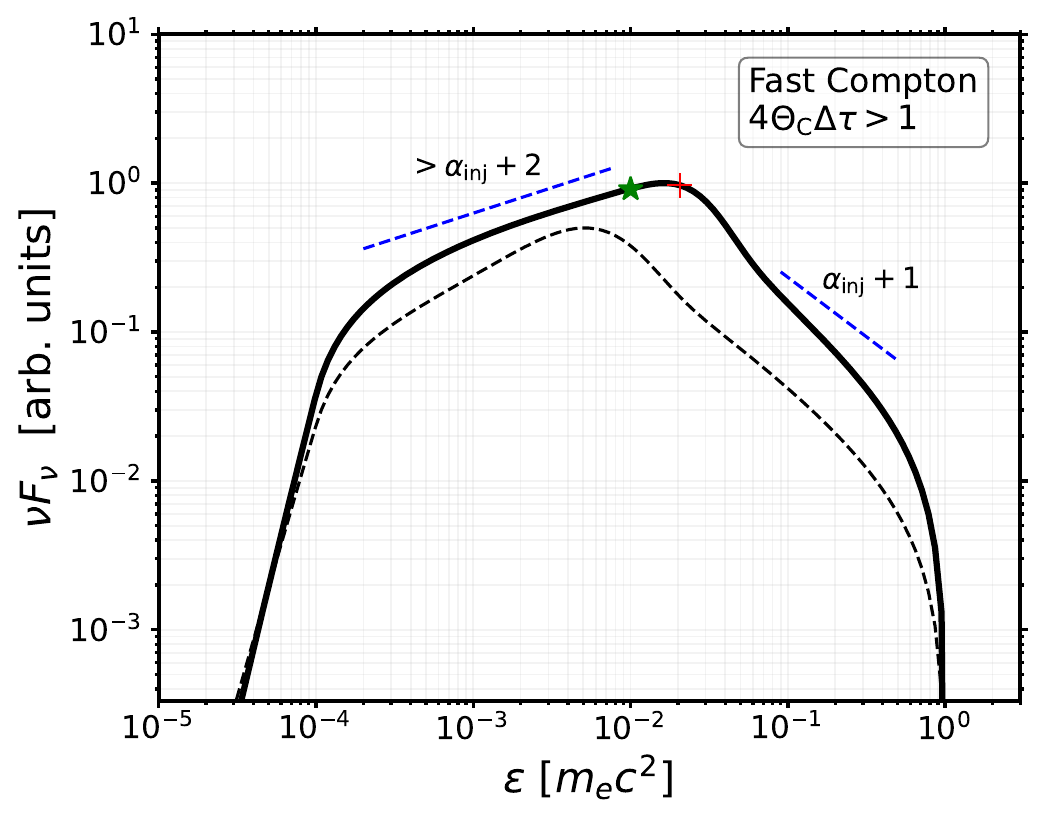}\\
    \caption{\textit{Left:} Schematics of the slow, marginally fast, and fast Compton regimes discussed in Section \ref{sec:fast_slow_Compton}. \textit{Right:} Simulation results in the three regimes. The red cross and the green star show the value of $4\Theta_{\rm C}$ and $1/\Delta \tau$ at $\tau_f$, respectively. The dashed blue lines show the theoretical slopes, which is determined solely by $\alpha_{\rm inj}$ in all instances except for the low-energy slope in the fast Compton case. In the simulations shown by the solid lines, the injection continues over a doubling of the radius. The black dashed lines have identical input parameters, except that the injection has been active for ten times longer. The solid lines are normalized to 1, while the dashed lines are normalized to 0.5 for clarity. 
    Parameter values are given in Table \ref{tab:values}.}
    \label{fig:Compton_regimes}
\end{figure*}

In Figure \ref{fig:Compton_regimes} on the left-hand side, we show schematics of the three spectra in the different regimes. Below $\epsilon_{\rm min}$, the dotted (dashed) line shows a Rayleigh-Jeans slope (Wien slope). Note that which spectral regime is the relevant one in a given context depends partly on $\Delta \tau$, but also on $\alpha_{\rm inj}$, since $\Theta_{\rm C}$ depends on the slope of the injected power-law spectrum.

\subsection{The influence of adiabatic cooling}\label{sec:adiabatic_cooling}
The discussion in Section \ref{sec:fast_slow_Compton} is valid in a plane parallel geometry and neglects the influence of adiabatic cooling on the spectrum. The characteristic timescale for adiabatic cooling corresponds to one doubling of the radius, which equals one halving of the optical depth. Thus, if a high-energy photon is injected at an optical depth $\tau_{\rm inj}$, it suffers Compton losses until its energy is $\epsilon_c = 2/\tau_{\rm inj}$ (assuming $2/\tau_{\rm inj} > 4\Theta_{\rm C})$. Thereafter, Compton losses are negligible with respect to adiabatic cooling. This leads to the power-law spectrum injected at $\tau_{\rm inj}$ having a high-energy cutoff at the photosphere of $\epsilon^{\rm cut}_{\rm ph} = \epsilon_c \, \frac{\phi(1)}{\phi(\tau_{\rm inj}/2)}$, which is approximately
\begin{equation}\label{eq:cutoff}
    \epsilon^{\rm cut}_{\rm ph} \sim  \left(\frac{\tau_{\rm inj}}{2}\right)^{-5/3}, \ \ 4\Theta_{\rm C} < \frac{2}{\tau_{\rm inj}}.
\end{equation}
Similarly, the Compton $y$-parameter for a low energy photon injected at $\tau_{\rm inj}$ should be evaluated as $y \sim 4\Theta_{\rm C}\frac{\tau_{\rm inj}}{2}$.

\subsection{Spectral shape as function of optical depth}\label{sec:spectral_shape_as_function_of_optical_depth}
The spectral shape as a function of optical depth depends on the nature of the injection. In this paper, we assume the photon number injection to be either constant as a function of radius, $dN_{\rm in}/dr = \rm{const.}$, or constant as a function of optical depth, $dN_{\rm in}/d\tau \propto r^2 dN_{\rm in}/dr = \rm{const.}$, where $N_{\rm in}$ is the number of injected photons (see equation \eqref{eq:source} below).

If the photon number injection is constant with radius, then at least half of the photons in the interaction region were injected during the last doubling of the radius, i.e., between $2\tau$ and $\tau$. Photons injected at $\tau_{\rm inj} \gg \tau$ are relatively few in number and contribute only with higher-order effects to the spectrum. The current spectral regime can be estimated by evaluating $4\Theta_{\rm C}(\tau) \Delta \tau$ and comparing it to unity, as in Section \ref{sec:fast_slow_Compton}. Note that $\Delta \tau = \tau$ in this case.

If instead the photon number injection is constant with optical depth, at least half of all photons were injected between $\tau_i$ and $\tau_i/2$. Thus, the overall spectral regime can be estimated by the value of $4\Theta_{\rm C}(\tau_i/2) \Delta \tau$, where $\Delta \tau = \tau_i/2$. If the photon distribution is in the slow or marginally fast Compton regime, i.e., $4\Theta_{\rm C}(\tau_i/2) \Delta \tau \lesssim 1$, then energy transfer via Compton scatterings has no large effect on the photon distribution at optical depths $\tau < \tau_i/2$. With few photons injected and little thermalization, the photon distribution is dominated by adiabatically cooling below $\tau_i/2$. Since the energy loss for each photon during adiabatic cooling is proportional to its current energy, the spectral shape (in a log-log plot) remains $\sim \,$constant. 
In the fast Compton regime, the spectral shape may continue to evolve, depending on the value of the Compton temperature at $\tau = \tau_i/2$.

%
\begin{table}[t]
    \centering
    \caption{Parameter values used for the right-hand panels in Figure \ref{fig:Compton_regimes}. All runs use $\epsilon_{\rm min} = 10^{-4}$ and $\epsilon_{\rm max} = 1$. The bottom row indicates the parameter value change used for the longer lasting runs, which are shown by a dashed lines in each panel}
    \begin{tabular}{llll}
    \hline
         Parameter & Slow Comp. & Marg. Comp. & Fast Comp. \\
    \hline
         $\tau_i$ & 200 & 40 & 200\\
         $\tau_f$ & 100 & 20 & 100 \\
         $\alpha_{\rm inj}$ & -2.3 & -1.9 & -1.8 \\
         $C({\bar r})$ & ${\bar r}^{-2}$ & 1 & 1 \\
         $--$ & $\tau_f = 10$ & $\tau_i = 400$ & $\tau_i = 2000$ \\
    \hline
    \end{tabular}
    \label{tab:values}
\end{table}
\subsection{Simulating the three regimes}\label{sec:simulating_three_regimes}
To verify the discussion in the previous sections, we make simulations in each of the three regimes. The spectra are obtained as follows.

The evolution of a photon distribution in interaction with thermal electrons is described by the Kompaneets equation. The Kompaneets equation in an ultra-relativistic outflow and modified to account for adiabatic cooling as the jet expands is given in \citet{VurmBeloborodov2016}. Neglecting induced emission and assuming constant $\Gamma$, it is given by \citep{Samuelsson2022}
\begin{equation}\label{eq:kompaneets_spherical}
\begin{split}
	&\frac{\partial}{\partial {\bar r}}\left({\bar r}^2 n_\epsilon \right) = s(\epsilon) \ + \\[2.3mm]
	&\frac{1}{\epsilon^2}\frac{\partial}{\partial \epsilon}{\Bigg [ }\frac{\epsilon^4}{{\bar r}^2}\left(\Theta_{\rm C} \frac{\partial \left({\bar r}^2 n_\epsilon \right)}{\partial \epsilon} + \left({\bar r}^2 n_\epsilon \right)\right) + \frac{2}{3} \frac{\epsilon^3\left({\bar r}^2 n_\epsilon \right)}{{\bar r}}{\Bigg ]}.
\end{split}
\end{equation}
\noindent Here, ${\bar r}$ is the normalized radius. ${\bar r} = r/R_{\rm ph}$, $n_\epsilon$ is the photon occupation number, related to the specific photon number density as ${\mathcal N}_\epsilon \propto \epsilon^2 n_\epsilon$, and $s(\epsilon)$ is a source term. The first term in brackets accounts for upscattering of low-energy photons while the second term accounts for downscattering of high-energy photons. The last term in brackets accounts for an ideal adiabatic cooling of the photon distribution as $\propto {\bar r}^{-2/3}$. The Kompaneets equation is solved using the solver from \citet{ChangCooper1970}. 

The source term injects a power-law distribution of photons as
\begin{equation}\label{eq:source}
\begin{split}
    s(\epsilon) \propto 
    \begin{cases}
        C({\bar r}) \epsilon^{p_{\rm inj}}, &\qquad \epsilon_{\rm min} < \epsilon < \epsilon_{\rm max},\\
        0, &\qquad \textrm{otherwise.}
    \end{cases}
\end{split}
\end{equation}
\noindent The source term is only non-zero at optical depths between $\tau_i$ and $\tau_f$. Note that $p_{\rm inj}$ is related to $\alpha_{\rm inj}$ as $\alpha_{\rm inj} = p_{\rm inj}+2$. The term $C({\bar r})$ determines the dependency on ${\bar r}$. If $C({\bar r}) = 1$, the injection is constant as a function of $r$. 
If instead, $C({\bar r}) = {\bar r}^{-2}$, the injection is constant as a function of $\tau$.

The Compton temperature is given by
\begin{equation}\label{eq:thetaC}
    4\Theta_\mathrm{C} = \frac{\int \epsilon^4 n \, d\epsilon}{\int \epsilon^3 n \, d\epsilon},
\end{equation}
\noindent obtained by solving equation \eqref{eq:kompaneets_spherical} in steady state without injection and adiabatic cooling. Since the Compton temperature is a function of the shape of the photon spectrum, it must be evaluated continuously. 

In Figure \ref{fig:Compton_regimes} on the right-hand side, we show the result of six simulation runs, whose parameters values are given in Table \ref{tab:values}. For all six simulations, the spectrum is evaluated just at the end of the injection zone at $\tau_f$. 

In each of the three panels, two simulation runs are shown. The solid lines show simulations that has run for one doubling of the radius. The red cross and the green star indicate the value of $4\Theta_{\rm C}$ and $1/\Delta \tau$, respectively, at the end of the simulation. It is clear that the simulated spectrum in each regime closely resembles the corresponding idealized schematic shown on the left. Adiabatic cooling is accounted for in the simulation runs but since the runs only last for one doubling of the radius, its effect is small. The dashed blue lines, with their respective slopes, are shown for clarity. In these cases, the relevant $\Delta \tau$ is $\tau_i - \tau_f$.

To study the effect of longer-lasting injection, we make three additional runs with identical parameter sets but where the injection has continued for ten times longer. These runs are shown by the black dashed lines in each panel. In each case, the two simulations have the same global characteristics, verifying the discussion in Section \ref{sec:spectral_shape_as_function_of_optical_depth}. However, the detailed spectral shapes are different. Specifically, the relations between the power-law slopes are modified. The relevant $\Delta \tau$ for the simulations with longer-lasting injection are $\Delta \tau = \tau_i/2$ for the slow Compton simulation (injection is proportional to $\tau$) and $\Delta \tau = \tau_f$ for the other two simulations (injection is proportional to $r$).

\section{Spectral evolution across the decoupling zone}\label{sec:evolution_across_decoupling_zone}
The decoupling zone is the region of optical depth wherein escaping photons contribute significantly to the observed fluence. Spectral change across the decoupling zone is, therefore, possible to detect in the observed time-resolved signal.\footnote{To observe such change requires that the pulse duration is longer than the time-resolution of the observing telescope.} The photon distribution can experience significant thermalization or additional dissipation across the decoupling zone. This leads to a three different scenarios.

\textit{i) No dissipation and no high-energy photons}. Only photons with energies above $2/\Delta \tau_{\rm dc}$ ($\sim 100~$keV) in the comoving frame have time to downscatter while traversing the decoupling zone (see \ref{sec:fast_slow_Compton} and \ref{sec:adiabatic_cooling}). If no such photons exist, and there is no additional dissipation, the comoving radiation is only affected by adiabatic cooling. Note that the Compton $y$-parameter for the low-energy photons is necessarily less than unity, since $4\Theta_{\rm C}$ cannot be larger than the maximum photon energy. The observer will see a spectral shape that is constant in time, but which decreases in energy continuously across the pulse. How large the total decrease is depends on the signal to background level. If we imagine the observer can distinguish the signal coming from optical depths of $\tau \sim 10$ until $\tau \sim 0.3$, then the observed spectrum would decrease in energy by a factor of $\sim \phi(10)/\phi(0.3) \sim 7$ across the pulse. 

\textit{ii) No dissipation but high-energy photons}.
In this case, the highest energy photons have time to downscatter. This leads to a spectral softening with time of the high-energy part of the observed spectrum, shifting the upper cutoff to lower energies. The downscattering is efficient until an optical depth $\tau = \tau_{\rm dc}/2\sim 5$, after which adiabatic cooling will dominate. The radiated energy as a function of $\tau$ peaks at $\tau_{\rm peak} \sim 3$ (see Figure \ref{fig:energy_vs_tau}). Thus, we expect a rapid softening of the high-energy part during the rise-time of the pulse, which stabilizes and remains $\sim\,$constant during the decay. The observed upper cutoff energy at the light curve peak can be estimated as
\begin{equation}\label{eq:high_energy_cutoff}
    E_{\rm cut}^{\rm obs} \sim \frac{2}{\Delta \tau_{\rm dc}} \left(\frac{\tau_{\rm dc}/2}{\tau_{\rm peak}}\right)^{-2/3}\frac{\Gamma m_ec^2}{1+z} \sim \frac{7~{\rm MeV} }{1+z} \ \Gamma_2,
\end{equation}
\noindent where $\Gamma_2 \equiv \Gamma/10^2$ and $z$ is the cosmological redshift. 

If the Compton temperature is exceptionally high, a slight hardening of the low-energy spectrum could be observed. In addition, we expect the previous behaviour of hard-to-soft evolution of the peak energy due to adiabatic cooling of the spectrum as a whole. This scenario is demonstrated by Run A in Section \ref{Sec:results}.

\textit{iii) Additional dissipation in the decoupling zone}.
In this case, the evolution of the observed spectrum will depend on the nature of the dissipation. If the injection is constant with radius, i.e., $C({\bar r}) = 1$ in the current framework, then the photon distribution will increasingly start to resemble the injected spectrum at smaller optical depths. This is because the total photon number in the interaction region doubles with each doubling of the radius, while the number of scatterings in between each doubling halves. At some point, the radiation is likely to enter the slow cooling Compton regime even if the injected power law slope is hard, and when $\tau <2/\epsilon_{\rm max}$, not even the highest energy photons have time to downscatter. This scenario is demonstrated by Run B in Section \ref{Sec:results} (although $\epsilon_{\rm max} =1$ in Run B so there are always some photons being downscattered).

If the dissipation is proportional to $\tau$, the magnitude of the spectral change in the observer frame depends on when injection started. If $\tau_i \gg \tau_{\rm dc}$, only a small fraction of the total photon number is injected in the decoupling zone and the situation will resemble \textit{i)} or \textit{ii)}.
Note that in the case of additional dissipation at low optical depths, the shape of the radiated energy as a function of $\tau$ given in Figure \ref{fig:energy_vs_tau} will be modified.



\section{Evolution of the observed spectrum}\label{Sec:results}
\begin{table}[t]
    \centering
    \caption{Parameter values used in Figure \ref{fig:spectral_evolution}. Both runs use $\epsilon_{\rm min} = 10^{-4}$ and $\epsilon_{\rm max} = 1$}
    \begin{tabular}{lll}
    \hline
         Parameter & Run A & Run B \\
    \hline
         $\tau_i$ & 50 & 50  \\
         $\tau_f$ & 10 & 3  \\
         $\alpha_{\rm inj}$ & -1.9 & -1.6 \\
         $C({\bar r})$ & ${\bar r}^{-2}$ & 1 \\
         $z$ & 2 & 0 \\
         $t_{\rm E}/t_{\rm dyn}$ & 1 & 1 \\
    \hline
    \end{tabular}
    \label{tab:values2}
\end{table}
In this section, we study the evolution of the observed signal in two simulated pulses, whose parameters are given in Table \ref{tab:values2}. The modeling of the comoving photon distribution is done in the same way as in Section \ref{sec:simulating_three_regimes}. In Run A, photon injection stops just at the beginning of the decoupling zone. Since $\epsilon_{\rm max}$ is high, Run A illustrates the case of \textit{ii) No dissipation but high-energy photons}, as discussed in Section \ref{sec:evolution_across_decoupling_zone}. In Run B, photon injection continues until $\tau = 3$. Thus, this run illustrates one possible scenario of \textit{iii) Additional dissipation in the decoupling zone}.

How we obtain the observed signal is outlined in Appendix \ref{sec:methodology}. In short, the observed signal is obtained using the code \texttt{Raylease}. We assume that the central engine emits a relativistic outflow during an active period between $t = 0~$s and $t = t_E$. A photon that experiences its last scattering at an optical depth $\tau$ and at an angle $\mu'$ reaches the observer at
\begin{equation}\label{eq:tobs}
    t_{\rm obs} = (1+z)\left[t_e + \frac{R_{\rm ph}}{c\beta \Gamma^2 \tau (1 + \beta \mu')} \right],
\end{equation}
\noindent where $t_e$ is the emission time from the central engine of the shell from which the photon decoupled. The observer time $t_{\rm obs}$ is given related to an imaginary observed photon, emitted at the line-of-sight, at $t_e=0$ and $r = 0$. 

The observed spectral flux at observer time $t_{\rm obs}$, $F_{\epsilon}(t_{\rm obs})$, is given by equation \eqref{eq:F_obs}. Since we consider a thin part of the jet with a width $\delta r \sim r/\Gamma^2$, it implies that $t_E$ is short. Therefore, it is safe to assume that $\Gamma$, the central engine luminosity, $L$, and the spectral shape are all independent of $t_e$. In this case, equation \eqref{eq:F_obs} simplifies to equation \eqref{eq:F_obs_simplified}, which is what we use in this paper. The photon distribution is tracked from $\tau_i$ until well above the photosphere, and 1000 snapshots of the comoving photon distribution at linear intervals of $\tau$ have been used to generate the observed time-resolved signal.

\begin{figure*}
\centering
    \includegraphics[width=0.49\linewidth]{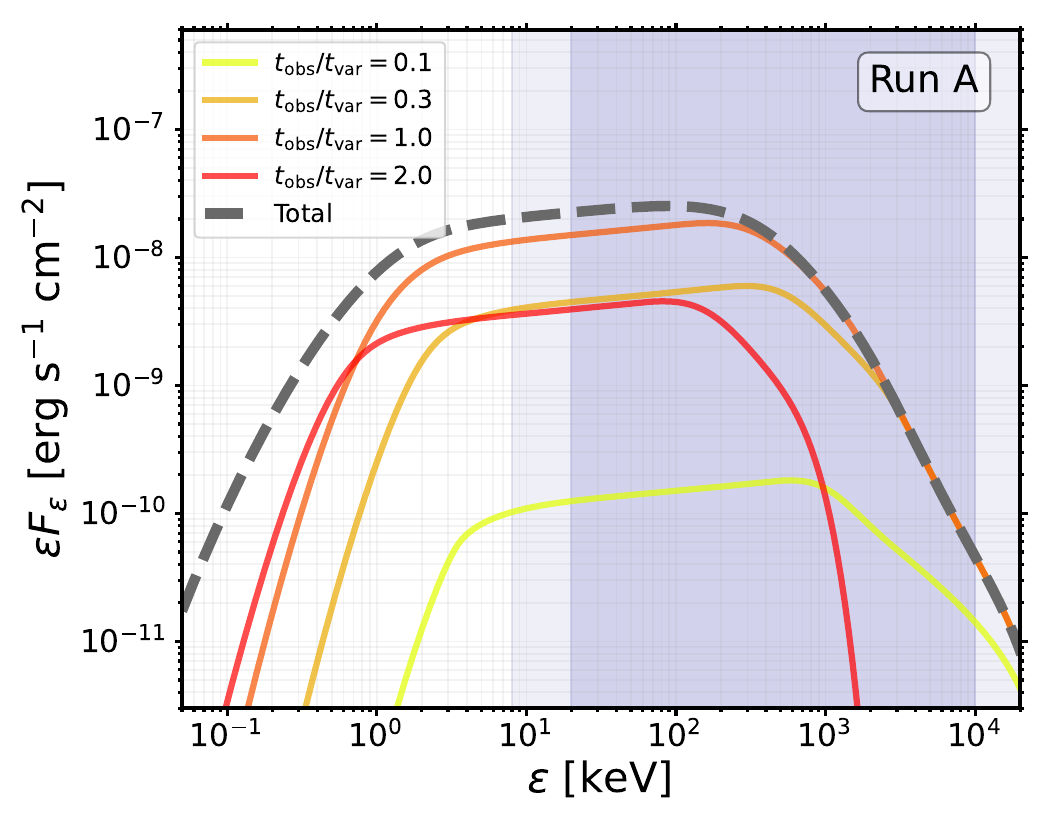}
    \includegraphics[width=0.49\linewidth]{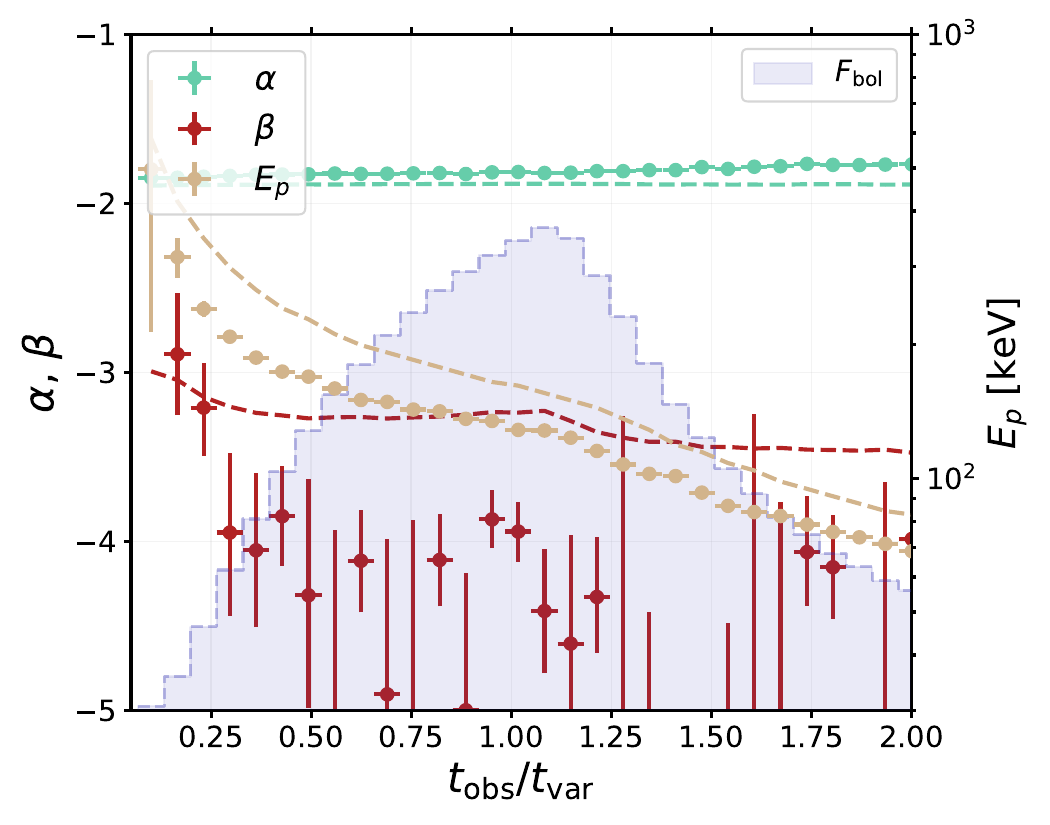}\\
    \includegraphics[width=0.49\linewidth]{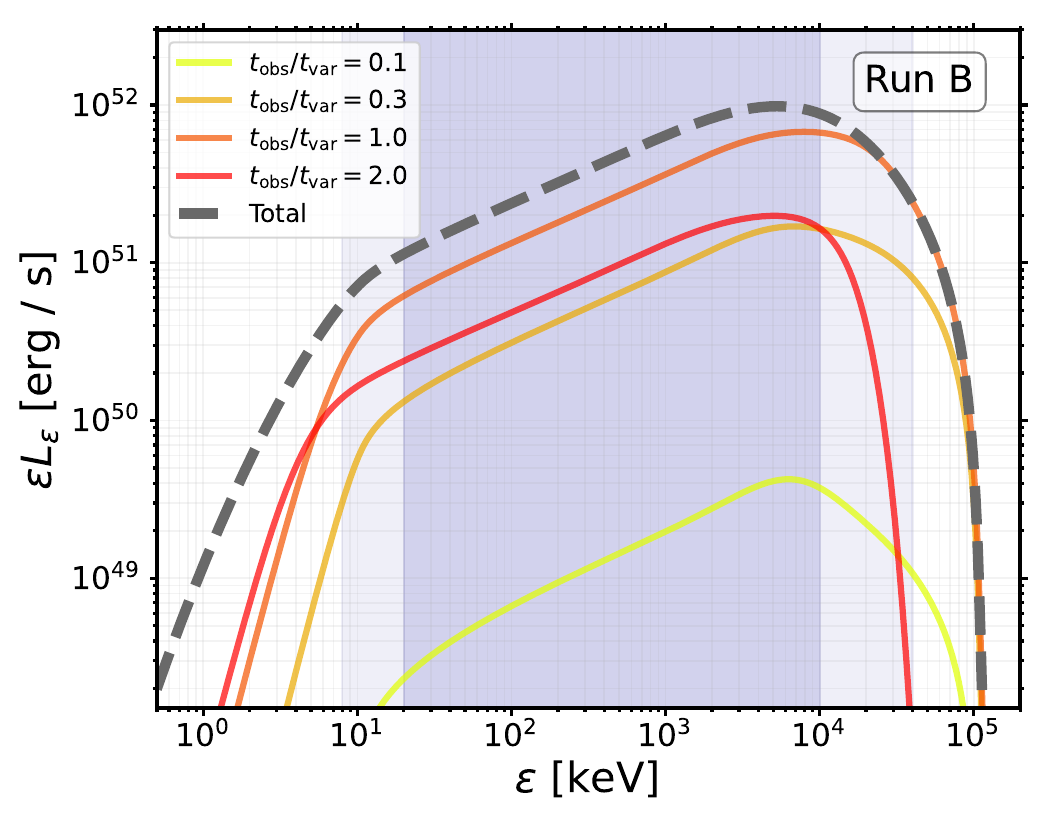}
    \includegraphics[width=0.49\linewidth]{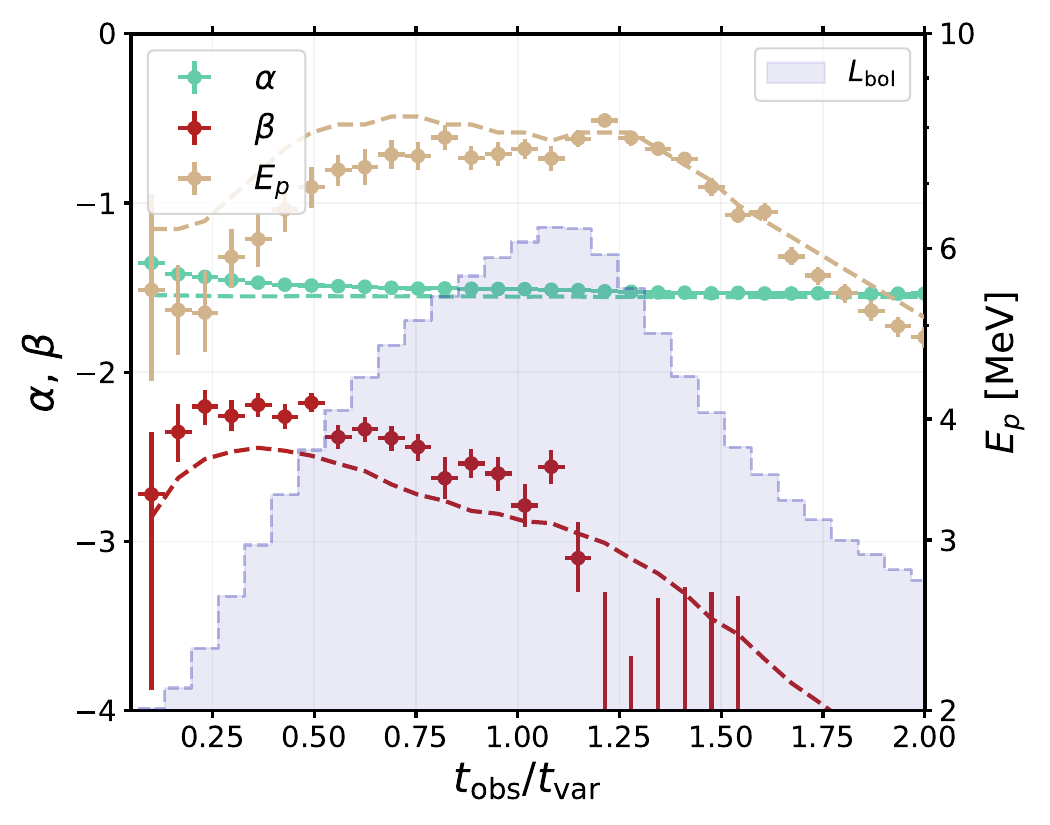}
    \caption{Observed spectral evolution for Run A (top) and Run B (bottom). \textit{Left:} Four snapshot spectra from early time (yellow) to late time (red). The grey dashed line shows the time-integrated spectrum. The purple shading shows the Fermi GBM energy window, with darker shading implying higher sensitivity. \textit{Right:} Evolution of the peak energy and the spectral indices as calculated theoretically (dashed lines) and as obtained by fitting a Band function to 30 mock data sets (points with error bars). Each mock data set has been forward-folded through the Fermi GBM response matrix, as explained in the main text. The flux in each bin is shown by the purple shading. In Run A, the fitted high-energy power law softens with time as comoving photons are downscattered. The peak energy decreases steadily due to adiabatic cooling. In Run B, the peak energy initially increases due to injection of high-energy photons at low optical depths. Parameter values are given in Table \ref{tab:values2}.}
    \label{fig:spectral_evolution}
\end{figure*}

\subsection{Intrapulse spectral evolution}
In the left panels in Figure \ref{fig:spectral_evolution}, four snapshot spectra are shown at different intervals of $t_{\rm obs}/t_{\rm var}$ for the two runs. Here, $t_{\rm var} = t_{\rm dyn}(1+z)$, and the dynamical time $t_{\rm dyn}$ is given by
\begin{equation}\label{eq:tdyn}
    t_{\rm dyn} = \frac{R_{\rm ph}}{2c\Gamma^2} = 0.2~{\rm s} \, L_{53} \, \Gamma_2^{-5},
\end{equation}
\noindent where $L_{53} \equiv L/10^{53}~$erg~s$^{-1}$. In both runs, we assume that radial spreading of the plasma causes a perceived jet active period of $t_E = t_{\rm dyn}$ (this assumption is discussed briefly in Section \ref{sec:assumptions}). This implies a constant width of the considered slice as $\delta r = R_{\rm ph}/2\Gamma^2$. The black dashed line in both panels shows the time-integrated spectrum, which is generated by first obtaining the spectral fluence via time integration of 100 snapshots of the observed spectrum, evenly log-spaced in time between $t_{\rm obs}/t_{\rm var} = 0.03$ and $t_{\rm obs}/t_{\rm var} = 50$, and then dividing the integrated fluence with $t_{\rm var}$. For Run A (top), the spectra are shown in the observer frame with an assumed redshift $z = 2$. In the case of Run B (bottom), the spectra are instead shown in the central engine frame with $L_\epsilon = F_\epsilon \times 4\pi d_l^2/(1+z)$, where $d_l$ is the luminosity distance.\footnote{The factor $(1+z)^{-1}$ accounts for the fact that we use the luminosity distance together with the spectral flux \citep[e.g.,][]{Hogg1999}.} The purple shading shows the energy sensitivity of the Fermi Gamma-ray Burst Monitor (GBM), with a darker color corresponding to higher sensitivity \citep{Meegan2009}. 


%
%

The first photons to reach the observer originate from the leading parts of the slice ($t_e = 0$) along the line-of-sight ($\mu'=1$). By comparison with equations \eqref{eq:tobs} and \eqref{eq:tdyn}, one obtains when $\Gamma \gg 1$ that photons observed at $t_{\rm obs}/t_{\rm var} = 0.1$ decoupled at $\tau = 10$, i.e., just at the beginning of the decoupling zone. For Run A, the injection is proportional to optical depth and the spectral regime can be evaluated by estimating $4\Theta_{\rm C}(\tau_i/2) \Delta \tau$, with $\Delta \tau = \tau_i/2$ (see Section \ref{sec:spectral_shape_as_function_of_optical_depth}). When the injection is proportional to radius, as in Run B, the spectral regime at $\tau = 10$ is instead governed by the value of $4\Theta_{\rm C}(\tau) \Delta \tau$ with $\Delta \tau = 10$. From the simulations, we find $4\Theta_{\rm C}(\tau_i/2)\Delta \tau = 0.7$ for Run A and $4\Theta_{\rm C}(\tau)\Delta \tau = 0.9$ for Run B. Thus, in both runs the radiation is in the marginally fast Compton regime and we expect the relation $\beta \approx \alpha - 1$ at early times.


At later times, the observed signal is a mix of the leading and trailing parts of the slice, emitting at lower and higher optical depths, respectively, as well as emission from various angles. For Run A, the downscattering of the high-energy radiation in the comoving frame leads to a softening of the high-energy spectrum with time. At the peak of the light curve, the spectrum exhibits a hardening at $\sim 2~$MeV. This corresponds to the cutoff energy estimated in equation \eqref{eq:high_energy_cutoff}. However, the contribution from trailing parts at higher optical depths transforms the cutoff in the comoving frame to a hardening in the observer frame. 

For Run B, the additional dissipation in the decoupling zone changes the spectral evolution. As more and more photons are injected at lower optical depths, the comoving photon distribution starts to increasingly resemble the injected distribution (see Section \ref{sec:evolution_across_decoupling_zone}). In the case of Run B, this leads to a rising peak energy, which only starts to decrease due to adiabatic cooling once injection ceases at $\tau = 3$. The high-energy power law also hardens and the spectral peak broadens during the light curve rise. 

Apart from the parameter values given in Table \ref{tab:values2}, the simulations assume $L = 10^{53}~$erg~s$^{-1}$, $\Gamma = 100$, $R_0 = 10^{10}~$cm, and $\Gamma_0 = 4$, where $R_0$ and $\Gamma_0$ are the radius and Lorentz factor at the base of the jet, respectively. Note that these four parameters only affect the absolute value of the luminosity and the variability time, none of which change the observed spectral shape as a function of $t_{\rm obs}/t_{\rm var}$.
\subsection{Peak energy and spectral indices}\label{sec:parameter_evolution}
To investigate the changing spectrum more quantitatively, in this section we study the evolution of the peak energy, $E_p$, the low-energy spectral index, $\alpha$, and the high-energy spectral index, $\beta$. We will calculate these quantities theoretically, as well as estimate what would be obtained in actual data.

The theoretical peak energy is calculated as the maximum of the $\epsilon F_\epsilon$ spectrum at each observer time. The spectral indices are subsequently obtained as
\begin{equation}
    \alpha, \beta = \frac{\log[N_\epsilon(\epsilon_{\alpha, \beta} + \delta \epsilon)] - \log[N_\epsilon(\epsilon_{\alpha, \beta})]}{\log(\epsilon_{\alpha, \beta}+\delta \epsilon) - \log(\epsilon_{\alpha, \beta})}.
\end{equation}
\noindent Here, $\epsilon_{\alpha}$ and $\epsilon_{\beta}$ are the energies at which the spectral indices $\alpha$ and $\beta$ are estimated, respectively, $\delta \epsilon$ is a small increment in energy, and $N_\epsilon = F_\epsilon/\epsilon$. The evolution of $\alpha$, $\beta$, and $E_p$ thus calculated are shown in the right-hand side of Figure \ref{fig:spectral_evolution} with dashed green, red, and yellow lines, respectively. For the figure, we use $\epsilon_{\beta} = 5E_p$ in Run A, $\epsilon_{\beta} = 3E_p$ in Run B, and $\epsilon_{\alpha} = E_p/10$ in both runs.

To obtain how the spectral evolution would be observed by a real telescope, we do as follows. First, we generate 30 spectra as a function of time for both Run A and Run B. To account for detector and background effects, we forward-fold each spectrum through the GBM response matrix. This implies that the fitting procedure can only account for the signal within the energy range of the GBM \citep[8~keV to 40~MeV,][]{Meegan2009}. For each spectrum, a mock data set is generated. The strength of the signal over the background in the mock data is determined by requiring a specific S/N at the peak of the light curve, $\mathcal{S}$. In other words, the mock data set in the brightest time bin have S/N~$=\mathcal{S}$. The S/N of all other time bins are given by the relative luminosity at that specific time. Each mock data set is subsequently fitted with a Band function \citep{Band1993}. The Band function describes a smoothly broken power-law photon spectrum using four parameters: $\alpha$, $\beta$, $E_p$, and an overall normalization. The analysis is performed in the Multi-Mission Maximum Likelihood (3ML) framework \citep{Vianello2015}.

The best-fit values obtained for $\alpha$, $\beta$, and $E_p$ from the Band fits in Run A and Run B are shown by points with error bars in Figure \ref{fig:spectral_evolution}. For the two figures, we used $\mathcal{S} = 1000$. The very high S/N was chosen so that the spectral evolution is more clearly visible. The error bars indicate $1\sigma$.

The evolution of the parameters confirms the discussion in the subsections above. 

\textit{High-energy spectral index, $\beta$}: In both cases, the early spectra have $\beta \approx \alpha - 1$. For Run A, the downscattering of photons in the comoving frame leads to a shortening of the high-energy power-law segment. Part of the power law persists until late times, as visible from the theoretically calculated value of $\beta$, which stays more or less constant. However, the fitted values of $\beta$ softens quickly during the pulse rise. The high S/N of the mock data set means that the cutoff is ``seen'' in the fit, making the fitted values of $\beta$ very low. At late times, the Band-function is not a good fit for the input spectrum, which is better described at high energies by a high-energy power law plus an exponential cutoff. 

In Run B, $\beta$ hardens initially due to the continued injection. The fitted value of $\beta$ remains high since the high-energy cutoff is outside of the GBM sensitivity window, and, thus, unaccounted for in the fit. 
After the peak of the pulse, most of the comoving radiation spectra have lost their high-energy tail. This leads to a rapid decrease of the flux above the peak energy in the observed spectrum. When the cutoff enters the GBM energy window at $t_{\rm obs}/t_{\rm var} \sim 1.2$, the fitted value of $\beta$ drops dramatically.

\textit{Peak energy, $E_p$}: Due to the smooth curvature of the Band function, together with the low value of $\alpha$, the fitted value of $E_p$ is always a factor of $\sim 2$ lower than that of the input spectrum in Run A. However, the general evolution is the same: with no injection of new photons in the decoupling zone, $E_p$ decreases continuously across the pulse. The rate at which the peak energy decreases stalls somewhat between $t_{\rm obs}/t_{\rm var} = 0.7$ and $t_{\rm obs}/t_{\rm var} = 1.2$. The duration of the stalling is related to $t_E$, with higher values of $t_E$ corresponding to longer-lasting stalling (this is true when the properties of the central engine are constant).

For Run B, the continued injection of photons in the decoupling zone means that $E_p$ increases from its initial value. This spectral change is rapid enough to overcome the decrease of the peak energy due to adiabatic cooling. The peak energy starts decreasing at $t_{\rm obs}/t_{\rm var} \sim 1.33$, which corresponds to line-of-sight emission from the trailing parts of the slice at $\tau_f = 3$. When injection stops, adiabatic cooling starts to dominate and the observed peak energy drops.

\textit{Low-energy spectral index, $\alpha$}: For both runs, the theoretical and fitted values of $\alpha$ stay roughly constant throughout the pulse. This implies that the Compton temperature in the decoupling zone is not high enough for any visible upscattering of the low-energy photons.

It is clear from the figure that the best fit Band parameters follows the theoretical estimates quite well, with the notable exception of $\beta$ in Run A. However, it is important to remember that the fitted values obtained depend on S/N, redshift, the energy window of the current detector, and the spectral model used in the fitting. Another key factor is that the duration of the pulse in the observer frame must be longer than the chosen temporal bin size. Otherwise, a time-resolved spectral analysis is impossible. 

\begin{figure}
\centering
    \includegraphics[width=\linewidth]{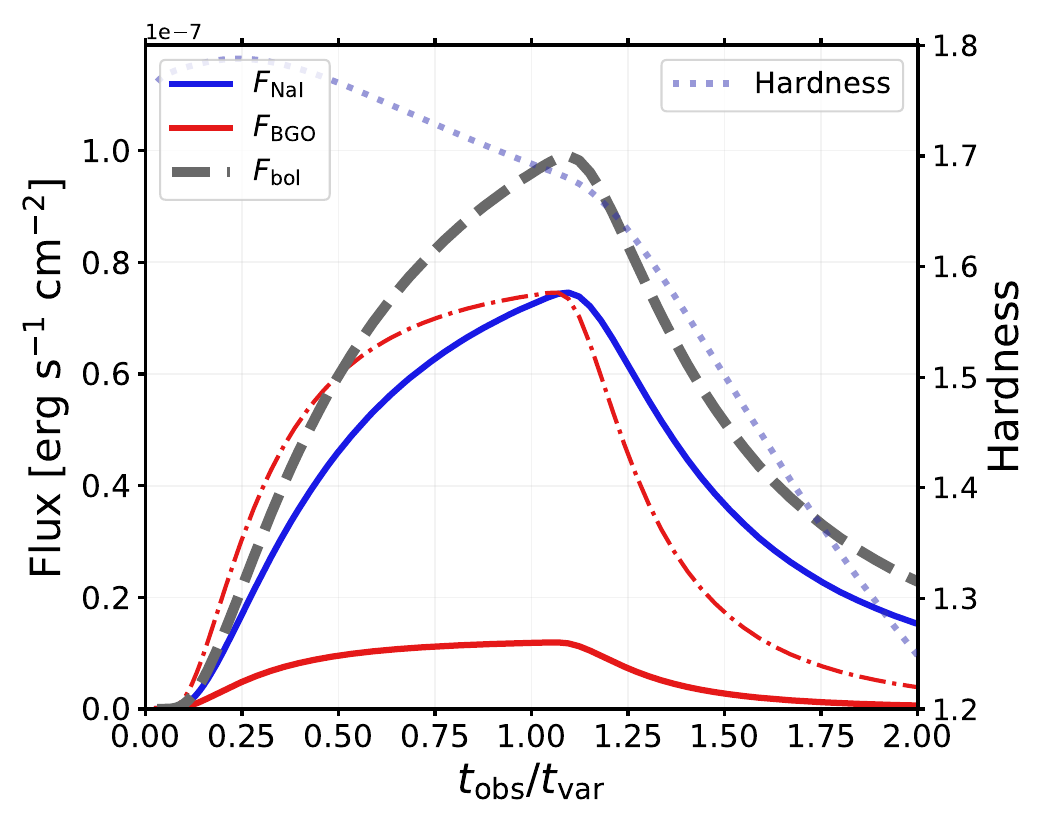}
    \includegraphics[width=\linewidth]{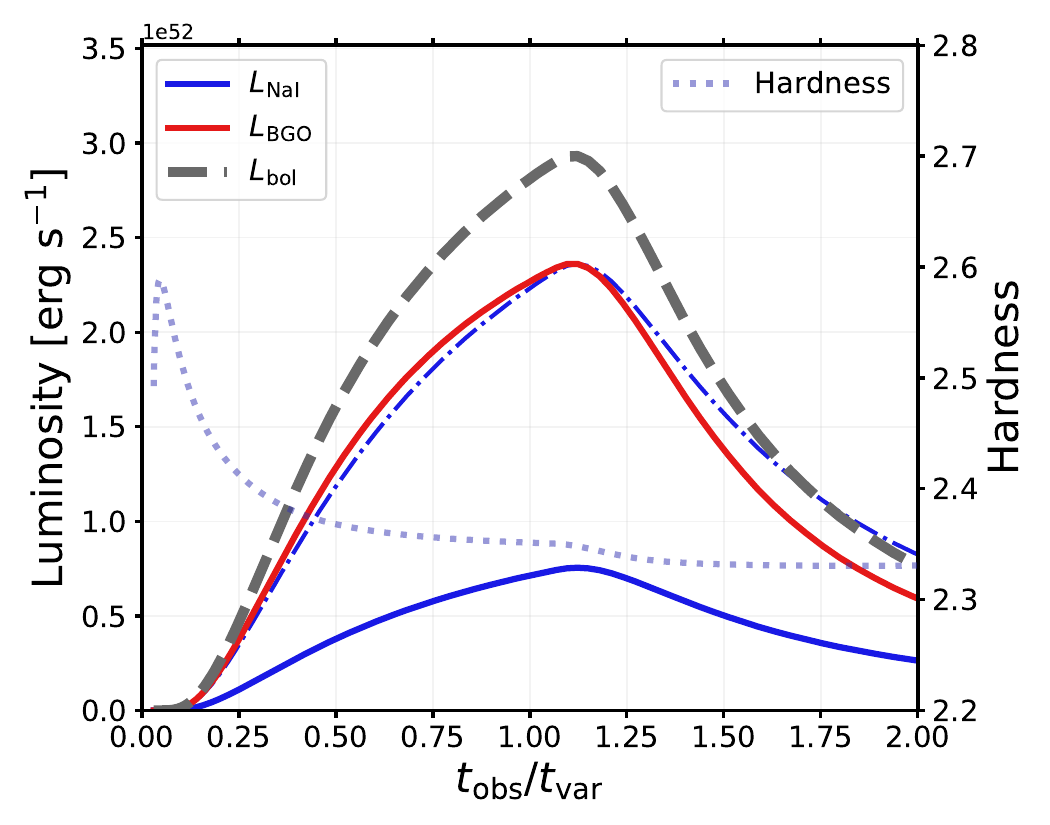}
    \caption{Observed light curve for Run A (top) and Run B (bottom). The dashed grey line shows the bolometric light curve, while the blue and red lines show the light curve in the NaI and BGO energy bands, respectively. The dotted-dashed line in the top (bottom) panel shows a scaled version of the light curve in the BGO (NaI) energy band to more clearly show the difference between the two light curves. Due to adiabatic cooling, the BGO light curve decreases more rapidly after the peak. The hardness as a function of time is shown on the right y-axis.}
    \label{fig:light_curve}
\end{figure}

\subsection{Light curve}\label{sec:light_curve}
In Figure \ref{fig:light_curve}, the observed light curves for the two pulses are shown. The grey dashed line shows the bolometric light curve, while the blue and red lines show the light curve in the Fermi NaI ($8-1000~$keV) and the Fermi BGO ($0.4-40~$MeV) energy ranges, respectively. The pulse profile is asymmetric. The duration of the rise time is related to the value of $t_E$, with smaller $t_E$ producing shorter rise times. Due to adiabatic cooling of the comoving radiation, the BGO light curve decreases quicker than the NaI light curve after the peak.

Apart from the light curve, we also plot the hardness as a function of time. The hardness is calculated as the ratio of the flux between $100-300~$keV to the flux between $50-100~$keV. The hardness evolution depends mostly on the position of the peak energy. In Run A, the hardness decreases continuously once the peak energy drops below $300~$keV at $t_{\rm obs}/t_{\rm var} \sim 0.25$. 
In Run B, $E_p$ remains much larger than $300~$keV throughout the pulse. Thus, the hardness remains constant after the first early evolution. 

\begin{figure}
\centering
    \includegraphics[width=\linewidth]{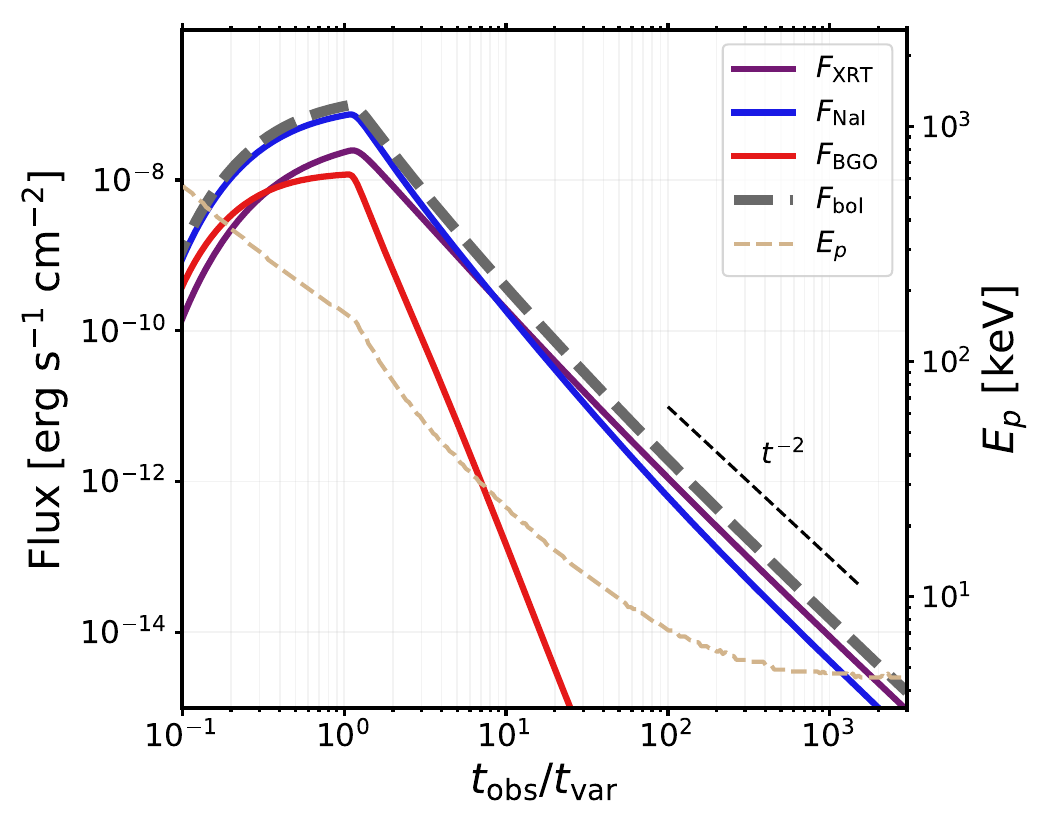}
    \caption{Very late-time light curve for Run A. Once $t_{\rm obs}/t_{\rm var}\gg1$, the bolometric flux decreases with time as $t^{-2}$. After the light curve peak, the peak energy first decreases roughly as $t^{-1}$. However, the decrease stalls around $t_{\rm obs}/t_{\rm var} \sim 100$. Note that due to the different scales on the y-axes, the grey dashed line indicating a $t^{-2}$ decay is valid only for the flux.}
    \label{fig:HLE}
\end{figure}

The light curve in Run A at very late times ($t_{\rm obs}/t_{\rm var}\gg1$) is shown in Figure \ref{fig:HLE}. In Figure \ref{fig:HLE}, the light curve in the \textit{Swift} XRT energy range \citep[$0.3-10~$keV,][]{Gehrels2004} is also shown. When $t_{\rm obs}/t_{\rm var}\gg1$, the bolometric flux decreases as $F_{\rm bol} \propto t^{-2}$ while the decrease of the peak energy stalls. These results are in line with those presented in \citet{PeerRyde2011}. The very late-time light curve for Run B is qualitatively similar. 

The very late-time behaviour of the pulse is interesting with regards to the early steep decay phase observed in X-rays in many GRBs \citep{Nousek2006}. The evolution of the luminosity and peak energy in Figure \ref{fig:HLE} are quite similar to those observed at the end of GRB pulses \citep{Ronchini2021,Tak2023}. However, given the small values of $t_{\rm var}$ for canonical GRB parameters, the early steep decay likely requires a different explanation in photospheric models of GRBs \citep{Peer2006ESD, Hascoet2012, Alamaa2024}, unless $\Gamma$ is small. 

\section{Discussion}\label{Sec:discussion}
\subsection{Assumptions}\label{sec:assumptions}
There are many simplifying assumptions in the current work. In this section, we list a few of them and discuss how they may effect our results.

\subsubsection{Power-law injected spectrum}\label{sec:power_law_injection}
Simulations of GRB jets find that a lot of energy is dissipated below the photosphere \citep{Lazzati2009, Gottlieb2019}. Several dissipation mechanisms are plausible and many of them generate a power-law distribution of photons, e.g., bulk Comptonization in a radiation-mediated shock \citep{Lundman2018, Ito2018, Samuelsson2022}, shear interactions in a structured jet \citep{Ito2013, VyasPeer2023}, pair cascades as a result of nuclear collisions \citep{Beloborodov2010}, and synchrotron emission from charged particles accelerated at a collisionless subshock \citep{LundmanBeloborodov2019, Levinson2020, Rudolph2023}. Thus, the choice of a power-law injection spectrum is motivated. 

However, the assumption of a power law that remains constant in time as the jet evolves is a strong one. 
If high-energy particles are present, as in the case of pair cascades and synchrotron emission, inverse Compton scattering on the charged particles can generate complicated radiation profiles \citep{Peer2005,Vurm2013}. Highly relativistic radiation-mediated shocks generate pairs and are subject to Klein-Nishina effects \citep{Lundman2018, Ito2018}. Furthermore, the injection is likely to evolve, specifically at low optical depths \citep{Levinson2012}.

In this paper, we have neglected a lot of physical complexity in favor of a very simple model that more clearly highlights the main points, which are the three spectral regimes at high optical depths and the evolving observed signal. According to the discussion in Section \ref{sec:spectral_shape_as_function_of_optical_depth}, the spectral regime of the radiation is roughly determined during one single doubling of the radius, across which the injection may be $\sim \,$constant. The two examples in Figure \ref{fig:spectral_evolution} clearly show that the observed signal evolves due to changes of the comoving radiation field. That the observed signal is affected by the changing local radiation in the jet is a general result. However, the details regarding how the observed signal is going to evolve will depend on the considered dissipation scenario. Thus, precise predictions for each case requires dedicated studies using physical models. 

\subsubsection{Other assumptions}\label{sec:other_assumptions}
The photon distribution is modeled using the Kompaneets equation \eqref{eq:kompaneets_spherical}. However, at small optical depths, the radiation field becomes increasingly anisotropic \citep{Beloborodov2011}. Equation \eqref{eq:kompaneets_spherical} then overestimates the energy transfer between photons and electrons \citep{VurmBeloborodov2016}, which leads to a quicker thermalization. If one accounts for this, the high-energy power law in Run A would survive for a longer time. 

The interaction region was assumed devoid of photons at optical depths $\tau>\tau_i$. As mentioned in Section \ref{sec:jet_profile}, this choice was inspired by the region downstream of a subphotospheric shock. However, in other dissipation scenarios, the energized photons may exist in the same local region as an initial thermal photon population at low temperature. The thermal population would decrease the Compton temperature in the interaction region, which in turn would increase the parameter space for the slow and the marginally fast Compton regimes. Such a situation may be interesting to investigate in a future work.

We ignored Klein-Nishina suppression of the cross-section. However, this does not significantly effect the result. The angle-integrated Klein-Nishina cross section is only lower by a factor $\sim 2$ compared to the Thomson cross section at $\epsilon = 1$. Including the Klein-Nishina corrections, the highest energy photons would arrive earlier.

We assumed a constant $\Gamma$ throughout the jet evolution. In reality, the high internal energy could lead to a re-acceleration of the plasma and a second fireball. If the outflow accelerates in the decoupling zone, the radiation would suffer from greater adiabatic losses. However, the increased Doppler boost would cancel out this effect in the observer frame \citep{Beloborodov2013} and the evolution of the peak energy would be the same. The observed light curve would be slightly different compared to that shown in Figure \ref{fig:light_curve}.

The spectra shown in Figure \ref{fig:Compton_regimes} are very ``clean'' compared to the complicated radiation profiles found in several other works \citep[e.g.,][]{Peer2005, Vurm2013, Ahlgren2015}. The main reason is the lack of relativistic electrons in the present work. Indeed, since the electrons are always kept at the Compton temperature, and we assumed $\epsilon_{\rm max} \leq 1$, the electrons are always nonrelativistic. That $\epsilon_{\rm max} \leq 1$ also ensures that there is no pair production. With relativistic electrons, the radiation profile becomes more complicated due to inverse Compton scattering creating bumps in the spectrum. However, relativistic charged particles are difficult to produce below the photosphere in GRBs \citep{LevinsonNakar2020} and simpler spectral shapes similar to the ones presented in Figure \ref{fig:Compton_regimes} are, therefore, expected.

We assumed $t_E = t_{\rm dyn}$. This is based on radial spreading of the plasma, which causes even an originally infinitesimally thin shell to have a width of $\delta r \sim R_{\rm ph}/\Gamma^2$ at the photosphere \citep[e.g.,][]{Peer2015}. With an assumption of constant $\Gamma$, this causes a perceived jet activity period of $t_E \sim t_{\rm dyn}$. However, $t_E$ can be both smaller and larger. If dissipation is due to a subphotospheric shock, compression of the plasma would lead to $t_E \sim t_{\rm dyn}/3$ \citep{SamuelssonRyde2023}. If the jet instead varies on timescales that are much longer than $t_{\rm dyn}$, the pulse duration would increase. 

Lastly, we assumed injection of high-energy photons up to $\epsilon_{\rm max} = 1$. If $\epsilon_{\rm max} \lesssim 0.1$, then no high-energy photons have time to downscatter in the decoupling zone. In this case, observed spectral evolution, apart from adiabatic cooling, requires additional dissipation in the decoupling zone.






\subsection{\texorpdfstring{$\alpha-\beta$}{alphabeta} relation}\label{sec:alpha_beta_relation}
The relation $\beta = \alpha - 1$ is a theoretical prediction that applies to radiation in the marginally fast Compton regime. Due to adiabatic cooling in the comoving frame and the observed signal being a superposition of many comoving spectra, this relation is going to be approximate in the observer frame, $\beta \approx \alpha - 1$. The best chance of detecting it is at early times, during the rise-time of the pulse.

As mentioned in Section \ref{sec:fast_slow_Compton}, the $\alpha-\beta$ relation is valid only when $\alpha_{\rm inj} < -1$. When $\alpha_{\rm inj} \gtrsim -1$ the Compton temperature is so high that the photon distribution in the interaction region quickly forms a pronounced Wien peak around $\Theta_{\rm C}$ (indeed $4\Theta_{\rm C} \approx \epsilon_{\rm max}/2$ for the injected photon distribution when $\alpha_{\rm inj} = -1$, see equation \eqref{eq:thetaC}). 

Thus, a clear prediction is that the $\alpha-\beta$ relation should only be observed when the detected low-energy slope is quite soft, $\alpha \lesssim -1$. Burst with harder $\alpha$ must be in the fast Compton regime, in which case we expect a cutoff above the peak.\footnote{This is true in the current framework and with the current model assumptions. It may not always be the case, see e.g., \citet{Beloborodov2010, VyasPeer2023}.} There could still be some signature at higher energies above the cutoff as in the bottom panels in Figure \ref{fig:Compton_regimes}. If such a feature is detected it would be a clear sign of Comptonization, as the spectral shape in the bottom panels of Figure \ref{fig:Compton_regimes} is quite complex and unique. 

On the other hand, when the injection is very soft, $\alpha_{\rm inj} < -2$, the $\nu F_\nu$-peak corresponds to $\epsilon_{\rm min}$ (top two panels in Figure \ref{fig:Compton_regimes}). In this scenario, one has an observed spectrum with a very hard $\alpha$, a low $E_p$, and a broken power-law at high energies, whose indices are related as $\beta_2 = \beta_1 - 1$. 

\subsection{Early-time spectra}\label{sec:difference_early_late}
During the pulse decay, emission at different optical depths and angles reach the observer simultaneously. In contrast, the first observed photons are all emitted within the same local region of the jet. The early-time spectrum is, therefore, the spectrum that most resembles the comoving one, Doppler boosted into the observer frame. 
As such, it should be similar to one of the typical spectra in the three Compton regimes in Figure \ref{fig:Compton_regimes}, given that dissipation has occurred at low optical depths. 

This part of the pulse can be very difficult to detect due to the low count rate. However, if detected, it can give information regarding the physical properties of the comoving radiation, the dynamics of the jet, and the details of subphotospheric dissipation. Furthermore, it allows for observation of the jet at a radius of up to one order of magnitude smaller the photospheric radius. 
\subsection{Pulse duration}\label{sec:pulse_duration}
The observed duration is highly sensitive to the value of $\Gamma$ (see equation \eqref{eq:tdyn}). The duration of the light curve shown for Run A in Figure \ref{fig:light_curve} is $\sim 1~$s for the chosen parameters. If $\Gamma$ is increased to 300, the observed duration shrinks to only $5~$ms and for $\Gamma = 50$, the duration is $\sim 40~$s. Since the Lorentz factor also scales the Doppler boost, pulses with shorter duration should have higher peak energies on average, scaling as $\delta t_{\rm obs} \propto E_p^{-5}$. 

\citet{Dereli2022} suggested that GRBs that exhibit an X-ray plateau may have low Lorentz factors, in the order of a few tens. Therefore, this sample may be interesting to investigate as the lower Lorentz can facilitate a time-resolved analysis.

\subsection{Dissipation radius and central engine activity}
Imagine a scenario where most of the subphotospheric energy dissipation takes place over a small distance, equal to one doubling of the radius or less. In this case, $\Delta \tau = \tau_i - \tau_f < \tau_f$. As long as $\alpha_{\rm inj}>-2$ and $\epsilon_{\rm max}> 1/\Delta \tau$, the comoving peak energy equals $\epsilon_{p} \approx \max[4\Theta_{\rm C}, 1/\Delta \tau]$ at the end of the dissipation zone (adiabatic cooling is small over one doubling of the radius). Since $1/\tau_f < 1/\Delta \tau$, all injected photons have time to be downscattered before the ejecta reaches the photosphere and the observed spectrum is well described by a cutoff power-law function.

As long as $4\Theta_{\rm C}$ is not much larger than $1/\Delta \tau$ at $\tau_f$, then $\tau_f$ can be estimated from the peak energy of the observed cutoff power-law spectrum as
\begin{equation}\label{eq:tau_f}
\begin{split}
    \tau_f &\sim 2\left[\frac{5\Gamma}{3(1+z)}\frac{m_e c^2}{E_p}\right]^{3/5} \\
    &\approx 110 \ \Gamma_2^{3/5} \, E_{p,2}^{-3/5}\,(1+z)^{-3/5},
\end{split}
\end{equation}
\noindent where we used equation \eqref{eq:cutoff}, a Doppler factor of $5\Gamma/3$ \citep{SamuelssonRyde2023}, and $E_{p,2} = E_p/(100~$keV). With a measured value of $t_{\rm var}$ from the light curve, the final dissipation radius can be estimated as $R_f = 2c\Gamma^2t_{\rm var}/(1+z)\tau_f$, where we used $R_f = R_{\rm ph}/\tau_f$ and equation \eqref{eq:tdyn}. 

If the dissipation is due to an internal collision the final dissipation radius is related to the central engine duration $\Delta t_{\rm eng}$ as $R_f \approx 2c\Gamma^2 \Delta t_{\rm eng}/\psi$, where $\psi$, equal to the ratio of the Lorentz factors of the fast and slow moving materials, is of the order a few.\footnote{This estimate assumes that the comoving densities in the fast and slow moving material are roughly equal \citep{Samuelsson2022}} Then, the central engine duration can be estimated from the observables as
\begin{equation}
    \Delta t_{\rm eng} \sim \frac{t_{\rm var}}{(1+z)}\frac{\psi}{\tau_f},
\end{equation}
\noindent where $\tau_f$ is given by equation \eqref{eq:tau_f}.

\subsection{Comparison to shock breakout emission}
Shock breakout is a likely explanation for the emission in low-luminosity GRBs and for the prompt emission in GRB 170817 \citep[e.g.,][]{Campana2006, Waxman2007, NakarSari2012, Gottlieb2018, Bromberg2018, Beloborodov2020}. 
\citet{LundmanBeloborodov2021} simulated the time-resolved observations from the shock breakout in GRB 170817 using a hydrodynamic code coupled to Monte-Carlo photons \citep[\radshock,][]{Lundman2018}. 
They also find a time-resolved spectrum whose shape evolves with time. However, the evolution is quite different compared to the two runs studied in this work. 

There are at least two reasons for the different evolution. Firstly, their injection is very hard, leading to $4\Theta_{\rm C} \lesssim \epsilon_{\rm max}$, in which case the high-energy signature is indistinguishable from the cutoff. Secondly, the dissipation is dynamic. The shock travels in a speed gradient within the dynamical neutron star ejecta and speeds up once it gets close to breakout, which modifies the dissipation. 
\section{Conclusion}\label{Sec:conclusion}
In this paper, we have studied how the evolution of the comoving radiation in the region of last scattering in an optically thick, relativistic jet effects the time-resolved observed spectrum. As photons leave the plasma over a range of optical depths (Figure \ref{fig:energy_vs_tau}), the average photon experiences multiple interactions within the decoupling zone. The photons that decouple at high optical depths are the first to reach the observer. We may therefore expect an evolving observed signal if dissipation or significant thermalization occurs in the decoupling zone. 

To model the energy dissipation, we assumed a simple injection of a power-law photon spectrum across a range of optical depth (the injection zone, see Figure \ref{fig:jet_scematic}). All photons were accumulated in an interaction region, where they were allowed to interact with the local plasma. No other photons apart from those injected existed in the interaction region. We assumed that the photons vastly outnumbered the electrons, such that the electrons were always kept in a thermal distribution at the Compton temperature, $\Theta_{\rm C}$. In this case, we found that the radiation at high optical depths exists in one of three different characteristic regimes, depending on the value of $4\Theta_{\rm C}\Delta \tau$, where $\Delta \tau$ is the duration of the injection.

In the slow Compton regime ($4\Theta_{\rm C}\Delta \tau < 1$) and the marginally fast Compton regime ($4\Theta_{\rm C}\Delta \tau \sim 1$), the spectrum consists of two power laws. Low-energy photons, $\epsilon < 1/\Delta \tau$, do not have time to be significantly effected by Compton scattering. Thus, the slope of the low-energy power law coincides with the slope of the injected spectrum, $\alpha_{\rm inj}$. In contrast, the high-energy photons lose energy efficiently to the electrons, which generates a high-energy power law with index $\alpha_{\rm inj} - 1$. In the fast Compton regime, photons pile up around $4\Theta_{\rm C}$. The different spectral shapes in the three regimes are given in Figure \ref{fig:Compton_regimes}.  

The photon distribution in the interaction region was tracked from the onset of injection to far beyond the photosphere, accounting for thermalization and adiabatic cooling. The observed signal as a function of time was obtained via equation \eqref{eq:F_obs_simplified}, which correctly accounts for the probability of emission at various optical depths and angles to the line-of-sight. We focused on a thin slice of the jet, whose emission created an observed pulse of typical duration $\delta t_{\rm obs} \sim t_{\rm var} = 0.2~\textrm{s} \, (1+z)L_{53}\Gamma_{2}^{-5}$. A complex GRB light curve would consist of many such subpulses. We assumed spherical symmetry and a bulk Lorentz factor $\Gamma \gg 1$.

The evolution of the observed signal across the pulse depends on the comoving photon distribution in the decoupling zone, and we classified three different scenarios. In the case of no dissipation and no high-energy photons (comoving energies $\gtrsim 100~$keV), the observed spectral shape is constant, but the whole spectrum decreases in energy by a factor of $\sim 7$ across the pulse due to adiabatic cooling. In the case of no dissipation but high-energy photons, the high-energy photons have time to downscatter within the decoupling zone. This leads to a spectral softening of the high-energy index, which is most pronounced during the rise-time of the pulse. This scenario is shown in the two top panels in Figure \ref{fig:spectral_evolution}. In the case of additional dissipation within the decoupling zone, the observed signal depends on the nature of the dissipation. One such example is shown in the two bottom panels in Figure \ref{fig:spectral_evolution}. 

Finally, we discussed the importance of early time spectra. The early spectra consists only of photons originating from the same local region within the jet and they are emitted at a fairly high optical depth, $\tau \sim 10$. If the comoving radiation is in the marginally fast Compton regime at this time, one obtains a relation between the high- and low-energy spectral indices as $\beta \approx \alpha - 1$. These early spectra requires bright GRBs to be seen. However, they would be highly illuminating if detected since they can give us valuable insights into the jet behavior before it becomes optically thin.\\[1.5mm]

I thank the anonymous reviewer for many valuable suggestions. F.A. is supported by the Swedish Research Council (Vetenskapsr\aa det, 2022-00347). I thank Fr\'ed\'eric Daigne, Robert Mochkovitch, Felix Ryde, and Christoffer Lundman for fruitful discussions. 
This research made use of the High Energy Astrophysics Science Archive Research Center (HEASARC) Online Service at the NASA/Goddard Space Flight Center (GSFC). In particular, I thank the GBM team for providing the tools and data.
\appendix
\section{Obtaining the observed signal}\label{sec:methodology}
To obtain the time-resolved, observed spectrum, two things are needed. The first is the probability distribution for any given photon to escape the outflow as a function of optical depth and angle. In a spherically symmetric, ultra-relativistic outflow, this is given by equation \eqref{eq:f}.
The second is the comoving specific photon number density at each point in space and time, ${\mathcal N}_{\epsilon}$. Under the assumption of spherical symmetry, the photon density is assumed to only vary with optical depth and time, i.e., ${\mathcal N}_{\epsilon} = {\mathcal N}_{\epsilon}(\tau, t)$. 

\subsection{Comoving photon distribution}\label{sec:mathcalN_inc_thermal}


If there is no dissipation and no high-energy photons in the decoupling zone, the photon distribution is only affected by adiabatic cooling. In this case, it is enough to know the spectral shape at a single optical depth and how the spectrum is affected by adiabatic cooling as a function of $\tau$ to obtain ${\mathcal N}_{\epsilon}$. 
This is the special case implicitly assumed in \citet{SamuelssonRyde2023}. In \citet{SamuelssonRyde2023}, the photon distribution as a function of optical depth was obtained using the spectrum at the photosphere, ${\mathcal N}_{\epsilon_{\rm ph}}$, together with the photon cooling function given in equation \eqref{eq:phi}. 

To obtain the photon distribution as a function of optical depth in the general case, including the effect of scatterings and potential dissipation, it is necessary to track ${\mathcal N}_\epsilon$ during the jet evolution. The modelling of the photon distribution is done as in Section \ref{sec:simulating_three_regimes}. The distribution is followed from the onset of the injection zone until well above the photosphere and snapshot spectra are saved at even intervals. 

Equation \eqref{eq:kompaneets_spherical} assumes an ideal adiabatic cooling $\propto~r^{-2/3}$. 
However, the cooling changes close to the photosphere once photons start to decouple. To account for this, we stop the idealized cooling at $\tau_{\rm cool} = 1.2^{3/2}$ \citep[c.f., equation \eqref{eq:phi} and discussion in][]{SamuelssonRyde2023}.
This prescription gives the correct average energy at the photosphere. However, it slightly overestimates the amount of cooling experiences when $\tau \gtrsim 1$ and slightly underestimates it when $\tau \lesssim 1$. To correct for this, the energy grid in each saved snapshot is shifted such that the mean photon energy ${\bar \epsilon}(\tau)$ is given by
\begin{equation}
\begin{split}
    {\bar \epsilon}(\tau) = 
    \begin{cases}
        {\bar \epsilon}^\star(\tau) \phi(\tau) & \qquad \rm{if} \ \tau \leq \tau_{\rm cool},\\
        {\bar \epsilon}^\star(\tau) \frac{\phi(\tau)}{(\tau/\tau_{\rm cool})^{2/3}} &\qquad \rm{if} \ \tau > \tau_{\rm cool},
    \end{cases}
\end{split}
\end{equation}
\noindent where ${\bar \epsilon}^\star(\tau)$ is the mean photon energy in the snapshot before the shift. This gives a very small correction to the observed signal. 

\subsection{Observed flux}\label{sec:observed_flux}

Consider a relativistic outflow consisting of many thin consecutive shells. We assume that each of the embedded shells acts independently and focus on one specific shell with a total number of photons, $N_\gamma$. The observed photon number originating from a volume element $dV$ relates to the emitted photon number as \citep{PeerRyde2011, Lundman2013}
\begin{equation}\label{eq:dN_obs}
    dN_{\rm obs} = \frac{(1+z)^2}{d_l^2}\frac{dN_{\rm em}}{d\Omega_v} = \frac{{\mathcal D}^2(1+z)^2}{d_l^2} N_\gamma f(V, \Omega_v')dV.
\end{equation}
\noindent Here, $d_l$ is the luminosity distance to the source, $dN_{\rm em}/d\Omega_v$ is the emitted photon number into solid angle $d\Omega_v$, ${\mathcal D} = [\Gamma(1-\beta\mu)]^{-1} = \Gamma(1+\beta\mu')$ is the Doppler factor, and $f(V, \Omega_v')$ is the probability function for arbitrary volume element and viewing angle. The term $(1+z)^2$ is the correction for redshift, which is necessary to include since we are using the luminosity distance but the equation is for the photon number. The term ${\mathcal D^2}$ in the second equality comes from the solid viewing angle transformation $d\Omega_v = d\Omega_v'/{\mathcal D}^2$. For a spherically symmetric outflow, this simplifies to 
\begin{equation}\label{eq:dN_obs_spherical}
    dN_{\rm obs} = \frac{N_\gamma(1+z)^2}{4\pi d_l^2} f(\tau, \mu') \, d\tau d\mu'.
\end{equation}
A photon that makes its last scattering at a time $t$, radius $r$, and angle $\mu$ compared to the radial direction, reaches the observer at a time ${\tilde t}_{\rm obs}/(1+z) = t + (D - r\mu)/c$. Here, $D$ is the distance to the source, which is slightly different compared to the luminosity distance $d_l$. Since the arrival times for all photons are delayed by a factor $(1+z)D/c$, we define a new observer time as $t_{\rm obs} = {\tilde t}_{\rm obs} - (1+z)D/c$. With this definition, a photon emitted at a time $t = 0~$s directly from the central engine ($r = 0~$cm) would reach the observer at $t_{\rm obs} = 0~$s \citep[such a photon is called the trigger photon in][]{Peer2008}. 

If we assume that all photons in the shell that decouple between an optical depth $\tau$ and $\tau + d\tau$ do so instantaneously when the shell reaches the optical depth $\tau$, then the time dependence can be described by a Dirac $\delta$-function. A shell that was launched from the central engine at a time $t_e$, and which travels with a velocity $v$, reaches an optical depth $\tau$ at time $t = t_e + R_{\rm ph}/\tau v$, where we used $\tau = R_{\rm ph}/r$. Using this, and the fact that $(1- \beta\mu) = [\Gamma^2(1+\beta\mu')]^{-1}$, one obtains
\begin{equation}\label{eq:dN_obs_dt}
\begin{split}
    dN_{\rm obs}(t_{\rm obs}) &= \frac{N_{\gamma}(1+z)}{4\pi d_l^2} f(r, \mu') \, d\tau d\mu' \times \\
    &\delta\left(t_e + \frac{R_{\rm ph}}{c\beta\Gamma^2\tau(1+\beta\mu')} - \frac{t_{\rm obs}}{1+z}\right).
\end{split}
\end{equation}
\noindent The total spectral flux in the observer frame from the shell, $F_\epsilon(t_{\rm obs})$, is obtained by multiplying the above formula with $\epsilon' {\mathcal N}_{\epsilon'}$ and integrating over the entire emitting volume. Here, $\epsilon' = (1+z)\epsilon/{\mathcal D}$ is the photon energy as measured in the comoving frame. Note that the term $N_\gamma f(r, \mu') \, d\tau d\mu$ already ensures a correct number of decoupling photons. Therefore, ${\mathcal N}_\epsilon$ should be a probability distribution, normalized as $\int {\mathcal N}_{\epsilon'} d\epsilon' = 1$. 

Putting it all together, and inserting $N_\gamma$ from Section \ref{App:N_bw}, we obtain the observed flux from a series of multiple consecutive shells emitted between $t_e = 0$~s and $t= t_E$, assuming spherical symmetry as 
\begin{equation}\label{eq:F_obs}
\begin{split}
    &F_\epsilon(t_{\rm obs}) = \frac{1+z}{4\pi d_l^2} \int_0^{t_E} dt_e \int_{-1}^1 d\mu' \int_0^\infty d\tau \, \frac{L}{4 kT'_0\Gamma_0}\, \times \\
    &f(\tau, \mu') \, \epsilon' {\mathcal N}_{\epsilon'} \, \delta\left(t_e + \frac{R_{\rm ph}}{c\beta\Gamma^2\tau(1+\beta\mu')} - \frac{t_{\rm obs}}{1+z}\right),
\end{split}
\end{equation}
\noindent where $k$ is the Boltzmann constant and $T'_0$ and $\Gamma_0$ are the temperature (given in equation \eqref{eq:T0}) and Lorentz factor at the base of the jet, respectively. 

\subsection{Steady source}
A detailed study regarding the time-resolved observed signal for different jet emission profiles, i.e., different functional shapes for e.g., $L(t_e)$ and $\Gamma(t_e)$, was performed by \citet{Meng2019}. Here, we make the simplifying assumption that the jet properties do not vary in time. We further assume that the steady jet properties lead to a comoving spectrum, ${\mathcal N}_\epsilon$, that is steady in time, such that ${\mathcal N}_\epsilon = {\mathcal N}_\epsilon(\tau)$.

When the central object is a steady source, the coefficient $L/4 kT'_0\Gamma_0$ in equation \eqref{eq:F_obs} can be evaluated outside the integral. Note that even in this case, one still needs to perform the integral over $dt_e$ because of its dependence in the $\delta$-function. Using the $\delta$-function to remove the integral over $d t_e$, one obtains
\begin{equation}\label{eq:F_obs_simplified}
\begin{split}
    F_\epsilon(t_{\rm obs}) &= \frac{1+z}{4\pi d_l^2} \frac{L}{4 kT'_0\Gamma_0} \, \int_{-1}^{1} d\mu' \int_{0}^{\infty} d\tau \, \times \\
    &f(\tau, \mu') \, \epsilon' {\mathcal N}_{\epsilon'} H(t_{e, \delta}) H(t_E - t_{e, \delta}),
\end{split}
\end{equation}
\noindent where $t_{e,\delta}$ is the value of $t_e$ that satisfies the $\delta$-function in equation \eqref{eq:F_obs}
\begin{equation}
    t_{e, \delta} = \frac{t_{\rm obs}}{1+z} - \frac{R_{\rm ph}}{c\beta\Gamma^2\tau(1+\beta\mu')}.
\end{equation}
\noindent In equation \eqref{eq:F_obs_simplified}, $H$ is the Heaviside step function, which implies that
\begin{equation}
\begin{split}
    H(t_{e, \delta}) H(t_E - t_{e, \delta}) = 
    \begin{cases}
        &1 \qquad {\rm if} \ 0~{\rm s} \leq t_{e, \delta} \leq t_E,\\
        &0 \qquad {\rm otherwise}.
    \end{cases}
\end{split}
\end{equation}
\noindent The Heaviside functions are included so that only regions emitted during the jet activity time contribute to the observed flux. Note that equation \eqref{eq:F_obs_simplified} is only valid when $L$ and $\Gamma$ are constant in time, otherwise there will be extra terms when the $\delta$-function is used to remove the integral over $t_e$. 

\subsection{\texttt{Raylease}}
\texttt{Raylease} is the name of the numerical code used to obtain the observed time-resolved signal. \texttt{Raylease} takes as input $L(t_e)$, $\Gamma(t_e)$, and ${\mathcal N}_{\epsilon}(\tau, t_e)$ and gives as output $F_\epsilon(t_{\rm obs})$ by solving equation \eqref{eq:F_obs}. One also need to specify $R_0$ and $\Gamma_0$, which in principle could be functions of $t_e$ as well. In the simplified case of constant jet properties considered in this paper, \texttt{Raylease} solves equation \eqref{eq:F_obs_simplified} for specified values of $L$, $\Gamma$, $R_0$, $\Gamma_0$, and ${\mathcal N}_{\epsilon}(\tau)$.
As the photon distribution is unspecified before injection begins in the current framework, the upper limit in the integral over optical depth in equation \eqref{eq:F_obs_simplified} is set to $\tau_i$.

Currently, \texttt{Raylease} assumes a spherically symmetric jet but it can quite easily be extended to include jet structure and an arbitrary viewing angle in the future.

\subsection{Photon number in infinitesimal shell}\label{App:N_bw}
This section derives the number of photons in each unit solid angle. The derivation assumes that the net diffusion between different solid angles is negligible. The subscript 0 used throughout this section implies that the quantity is measured at the base of the jet, where the internal pressure starts efficient bulk acceleration. For the derivation in this section, the Lorentz factor $\Gamma$ is allowed to vary with radius.

The number of photons per unit solid angle in an infinitesimally thin shell is given by 
\begin{equation}
    \frac{dN_\gamma}{d \Omega} = \frac{d{\dot N}_\gamma}{{d \Omega}} dt_e.
\end{equation}
\noindent Assuming a negligible photon production once bulk acceleration of the outflow begins (see below), the number of photons follows conservation of particle number as $dN_\gamma = n'_\gamma r^2 \Gamma d\Omega dr$, which gives $d{\dot N_\gamma}/d \Omega = n'_\gamma r^2 \Gamma \beta c$, where $\beta c = dr/dt$. Similarly, we obtain $d {\dot M}/d\Omega = m_p n'_p r^2 \Gamma \beta c$. 

Define the specific entropy per baryon, $\eta(\Omega)$, such that $dL/d\Omega = \eta(\Omega)c^2d{\dot M}/d\Omega$, where $dL/d\Omega$ is the outflow power per unit solid angle. The specific enthalpy, $h'$, and the Lorentz factor, $\Gamma$, both of which varies with radius, has a product that is independent of radius as $h'\Gamma = \eta$ \citep[e.g.,][]{LevinsonNakar2020}. At the base of the jet, the Lorentz factor is small and the entropy is dominated by the pressure from the radiation and the pairs, $h'_0 \approx 4p_0/n'_0m_pc^2$, where $p_0$ is the total pressure and $n'_0$ is the baryon number density. Assuming a thermal equilibrium between the pairs and the radiation such that they can be described by a common temperature $T'_0$, the pressure is given by $p_0 \approx n'_{q,0}k T'_0$, where $n'_{q,0} = n'_{\gamma,0} + 2n'_{\pm,0}$ is the photon plus pair density. As the temperature drops, the pairs all recombine until $n'_q = n'_\gamma$. Assuming that recombination is the only source of photon production after the early hot phase, the ratio $n'_q/n'_p$ is conserved along streamlines and equal to $n'_{q,0}/n'_{p,0}$. Hence, we get
\begin{equation}
    n'_\gamma = n'_p \frac{\eta m_pc^2}{4kT'_0 \Gamma_0} = n'_p\frac{(dL/d\Omega) m_p}{(d{\dot M}/d\Omega)4kT'_0 \Gamma_0}
    = \frac{(dL/d\Omega)}{4kT'_0 \Gamma_0 \, r^2 \Gamma \beta c},
\end{equation}
\noindent which gives 
\begin{equation}
    \frac{dN_\gamma}{d \Omega} = \frac{dL}{{d \Omega}} \frac{1}{4kT'_0\Gamma_0}dt_e.
\end{equation}
\noindent Given that pairs and radiation are in thermodynamic equilibrium at the base of the jet, $T'_0(\Omega)$ is given by \citep{LevinsonNakar2020}
\begin{equation}\label{eq:T0}
    T'_0(\Omega) = \left[ \frac{dL}{d\Omega} \frac{3}{11}\frac{1}{R_0^2ac\Gamma_0^2\beta_0}\right]^{1/4}.
\end{equation}

\bibliographystyle{mnras}
\bibliography{References}
\end{document}